\documentclass[review]{elsarticle}

\journal{Artificial Intelligence}

\usepackage{microtype}
\usepackage{graphicx}
\usepackage{subfigure}
\usepackage{booktabs} % for professional tables
\usepackage{amsmath}
\usepackage{amsthm}
\usepackage{amssymb}
\usepackage{algorithm}
\usepackage{algorithmic}
\usepackage[dvipsnames]{xcolor}

\newcommand\numberthis{\addtocounter{equation}{1}\tag{\theequation}}
\renewcommand{\bar}{\overline}

\newcommand{\cE}{\mathcal{E}}
\newcommand{\cF}{\mathcal{F}}

\newcommand{\cN}{\mathcal{N}}

\newcommand{\cS}{\mathcal{S}}

\newcommand{\C}{\mathbb{C}}
\newcommand{\R}{\mathbb{R}}

\newcommand{\h}{{\rm I}\kern-0.18em{\rm H}}

\newcommand{\p}{{\rm I}\kern-0.18em{\rm P}}
\renewcommand{\P}{{\rm I}\kern-0.18em{\rm P}}
\newcommand{\E}{{\rm I}\kern-0.18em{\rm E}}
\newcommand{\Z}{{\rm Z}\kern-0.18em{\rm Z}}
\newcommand{\1}{{\rm 1}\kern-0.24em{\rm I}}
\newcommand{\N}{{\rm I}\kern-0.18em{\rm N}}

\newcommand{\x}{\mathbf{x}}

\newcommand{\argmin}{\mathop{\mathrm{argmin}}}

\newcommand{\norm}[1]{\left\|#1\right\|}

\newcommand{\subG}{\mathsf{subG}}

\renewcommand{\hat}{\widehat}
\newcommand{\tr}{\textrm{tr}}
\renewcommand{\bar}{\overline}

\newcommand{\w}{\mathbf{w}}
\newcommand{\y}{\mathbf{y}}
\renewcommand{\u}{\mathbf{u}}
\newcommand{\boldeta}{\boldsymbol{\eta}}

\newtheorem{lemma}{Lemma}
\newtheorem{proposition}{Proposition}
\newtheorem{theorem}{Theorem}
\newtheorem{definition}{Definition}
\newtheorem{example}{Example}
\newtheorem{remark}{Remark}
\newtheorem{assumption}{Assumption}

\newcommand{\newreptheorem}[2]{%
\newenvironment{rep#1}[1]{%
 \def\rep@title{#2 \ref{##1}}%
 \begin{rep@theorem}}%
 {\end{rep@theorem}}}
\makeatother

\newreptheorem{theorem}{Theorem}
\newreptheorem{proposition}{Proposition}

\begin{document}

\begin{frontmatter}

\title{Accurate Parameter Estimation for Risk-aware Autonomous Systems}

\author{Arnab Sarker\corref{mycorrespondingauthor}\fnref{mit}}
\cortext[mycorrespondingauthor]{Corresponding author}
\ead{arnabs@mit.edu}
%\address{Cambridge, MA}
\author{Peter Fisher\fnref{mit}}
\author{Joseph E. Gaudio\fnref{mit}}
\author{Anuradha M. Annaswamy\fnref{mit}}
\fntext[mit]{Massachusetts Institute of Technology}
\begin{abstract}
Analysis and synthesis of safety-critical autonomous systems are carried out using models which are often dynamic. 
Two central features of these dynamic systems are parameters and unmodeled dynamics. 
Much of feedback control design is parametric in nature and as such, accurate and fast estimation of the parameters in the modeled part of the dynamic system is a crucial property for designing risk-aware autonomous systems. 
This paper addresses the use of a spectral lines-based approach for estimating parameters of the dynamic model of an autonomous system. 
Existing literature has treated all unmodeled components of the dynamic system as sub-Gaussian noise and  proposed parameter estimation using Gaussian noise-based exogenous signals. 
In contrast, we allow the unmodeled part to have deterministic unmodeled dynamics, which are almost always present in physical systems, in addition to sub-Gaussian noise. 
In addition, we propose a deterministic construction of the exogenous signal in order to carry out parameter estimation.  
We introduce a new tool kit which employs the theory of spectral lines, retains the stochastic
setting, and leads to non-asymptotic bounds on the parameter estimation error.
Unlike the existing stochastic approach, these bounds are tunable through an optimal choice of the spectrum of the exogenous signal leading to accurate parameter estimation. 
We also show that this estimation is robust to unmodeled dynamics, a property that is not assured by the existing approach. 
Finally, we show that under ideal conditions with no unmodeled dynamics, the proposed approach can ensure a $\tilde{O}(\sqrt{T})$ regret, matching existing literature. 
Experiments are provided to support all theoretical derivations, which show that the spectral lines-based approach
outperforms the Gaussian noise-based method when unmodeled dynamics are
present, in terms of both parameter estimation error and Regret obtained using the parameter estimates with a Linear Quadratic Regulator in feedback.
\end{abstract}


\end{frontmatter}

% \linenumbers

\section{Introduction}
\label{s:Introduction}
From space to healthcare, energy to transportation, manufacturing to agriculture, autonomous systems are gathering increased attention and importance. The synthesis and analysis of autonomous systems pose formidable challenges to overcome which a systematic theory of risk-aware autonomous systems is necessary. A crucial building block of this theory forms the focus of this paper, and pertains to a new tool for designing risk-aware algorithms for use in autonomous systems.  

Safety-critical autonomous systems abound in a variety of sectors including transportation, energy, robotics, and manufacturing. A central property of these systems is self-governance in the presence of abrupt changes in the system environment. Many of these abrupt changes can be modeled in the form of parametric changes in a dynamic system. The challenge is to minimize the risk that occurs in this context using real-time tools. The new tool that we propose is paper is the use of sub-Gaussian spectral lines for parameter estimation. We will show in this paper that this tool (a)  can be employed in finite-time, (b) has the ability to accurately estimate the unknown parameters, and (c) enables minimization of risk of failure during online performance of the autonomous system.

The specific class of dynamic systems we will focus on is a linear time-invariant systems whose parameters are to be estimated, a problem that has been investigated at length in recent years \cite{Abbasi2011, Dean_2018a, Dean_2018, Foster_2020, Matni2019, Sarkar_2019, Sarkar_2019a, Simchowitz_2019, Simchowitz_2018, Tsiamis_2019, Wagenmaker2020}.  In contrast to the approaches taken in these papers where the exogenous input is chosen using independent realizations of Gaussian noise, we choose a deterministic signal constructed using the theory of spectral lines \cite{Boyd_1986,Boyd_1983}, which has been demonstrated to be effective in learning the parameters of the system, c.f. \cite{Anderson_1977,Anderson_1982,Narendra_1987,Willems2005}.

\begin{figure}
    \centering
    \includegraphics[width = 0.8\textwidth]{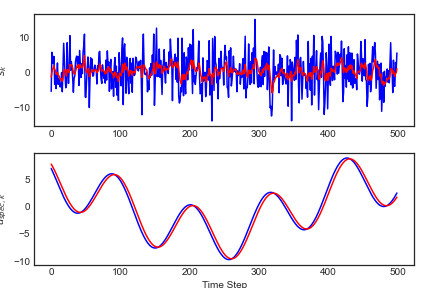}
    \caption{Examples of inputs dictated to a physical system (blue) and their realized counterparts (red), represented by feeding the planned inputs through a low pass filter to simulate physical constraints on an actuator. While the frequency content of the spectral-lines based signal (bottom) is unchanged by the low pass filter, the realized white noise signal (top) no longer consists of independent Gaussian noise.}
    \label{fig:inputs}
\end{figure}

While randomization of the control input allows for clean analysis of estimation using well-known tools, it is undesirable for several reasons in safety-critical autonomous systems. First and foremost, in safety-critical systems, the addition of intentional randomness of exogenous inputs is undesirable as it may excite unmodeled dynamics \cite{Dydek2008, Gorinevsky2012}.  Such unmodeled dynamics are almost always present due to several reasons in any physical system \cite{DoyleStein81}. No dynamic system can be perfectly modeled; modeling errors due to nonlinearities, higher-order linear dynamics, and/or unknown underlying mechanisms are inevitable. Second, any exogenous input that is synthesized using a computational algorithm has to be physically realizable.  Typically actuators are the entities in an autonomous system that perform this role of transduction from a digital signal to a physical signal in the appropriate energetic domain (ex. position, torque, heat, current). All actuators have limited bandwidth  and therefore poor fidelity in reproducing high frequencies (see Figure \ref{fig:inputs}).  Finally, Gaussian noise in principle is unbounded. This poses two obvious problems in light of the two issues mentioned above – they can excite any unmodeled dynamics thereby introducing significant inaccuracies in the parameter estimation which in turn can pose a risk for any estimation-based closed-loop control decisions. It may not be simply possible to accurately realize the prescribed input $u_k$ \cite{Lavretsky2013} thereby introducing additional inaccuracies in the estimation. 

This paper proposes a new tool for designing risk-aware algorithms for use in autonomous systems, in the form of sub-gaussian spectral lines. This represents a finite-time tool and  leads to estimates of the unknown parameter in the underlying dynamic model of the autonomous system. A deterministic choice of the exogenous signal is proposed, and constructed using a combination of spectral lines. We show that using this tool, a tunable parameter error bound can be determined. We also show that reduced regret can be achieved, and therefore proactively quantify and mitigate risks against task goals and safety. Formal guarantees both on the parameter error bounds and on the risk of failure are provided.

\section{Problem Formulation}

This work is motivated by physical systems that are governed by non-linear dynamics and perturbed by unobserved noise, where observations are made at discrete time steps. 
That is, the general system-model which is considered is of the form
\begin{equation} \label{eq:intro-nonlinear-general}
    x_{k+1} = f(x_k, u_k) + \eta_k\,, \quad \eta_k \stackrel{i.i.d.}{\sim} \subG(\sigma^2 I) \,,
\end{equation}
where $x_k \in \R^n$ represents the state of a dynamical system, $u_k \in \R^{m}$ represents an exogenous input, and $\eta_k \in \R^n$ represents unobserved disturbances. 
In practice, the system above can have multiple equilibrium points, and a practitioner may have the objective of steering the system towards a particular equilibrium point. 
In a region around the equilibrium point, one would expect the dynamics of the system to be dominated by linear dynamics, and it is reasonable to model the system as being a linear system in the presence of unmodeled dynamics.
Hence, the starting point for this paper is a system-model of the form
\begin{align}
x_{k+1} &= A_* x_k + B_* u_k + w_k+ \eta_k,\label{unmodeled-intro-x} \\
&\eta_k \stackrel{i.i.d.}{\sim} \subG(\sigma^2 I)\,, \label{unmodeled-intro-eta} \\
w_{k} &= f(x_0, w_{k-1}, \dots, w_0, u_k, \dots u_0 ) \label{unmodeled-intro-w}\,.
\end{align}
In this setting, the observations $x_k$ are measured, $A_*$ and $B_*$ are fixed but unknown, $\eta_k$ is unobserved, and the control input $u_k$ is selected by the practitioner. 
The values $w_k \in \R^n$ then represents the effect of all unmodeled components in the system and is therefore unknown. 
In practice, the presence of $w_k$ may be due to higher-order linear dynamics, nonlinearities, and/or due to other effects that cannot be precisely accounted for \cite{DoyleStein81, Khalil2002}. 
Typically $w_k$ remains small for low amplitudes and/or low frequencies and large if either amplitudes or frequencies become large. 
It is therefore often desired to keep the frequency content of $u_k$ small (see descriptions of $l_m(\omega)$ in \cite{DoyleStein81}) and amplitude small \cite{Khalil2002}. 
The unmodeled dynamics in \eqref{unmodeled-intro-w} could be even more complex  with dependence on $x_k$ for all $k$. 
Together with $w_k$ and $\eta_k$, we note that the actual state has unmodeled components both due to deterministic mechanisms due to \eqref{unmodeled-intro-w} and stochastic mechanisms represented in \eqref{unmodeled-intro-eta}.

In contrast to the above representation of a model of a safety-crticial system, the machine learning community has focused on modeling of a linear dynamical system and learning its parameters using independent Gaussian noise as the exogenous input \cite{Abbasi2011, Dean_2018a, Dean_2018, Foster_2020, Matni2019, Sarkar_2019, Sarkar_2019a, Simchowitz_2019, Simchowitz_2018, Tsiamis_2019, Wagenmaker2020}.  
Specifically, these works consider a system-model of the form
\begin{equation}
\label{lti-noisy-intro}
x_{k+1} = A_* x_k + B_* u_k + \eta_k, \quad \eta_k \stackrel{i.i.d.}{\sim} \subG(\sigma^2 I) \,.
\end{equation}
In the machine learning literature, the control input $u_k \in \R^m$ is either selected solely according to independent realizations of $\cN(0, \sigma_u^2 I)$, or is perturbed by independent realizations of Gaussian noise. More importantly, the underlying analysis in these papers neglects the presence of $w_k$ as in \eqref{unmodeled-intro-x}-\eqref{unmodeled-intro-w} altogether. 

As discussed above, while randomization of the control input allows for clean analysis of estimation using well-known tools, it is undesirable for several reasons. 
In safety-critical systems, the addition of intentional randomness of the exogenous input is undesirable as it may excite unmodeled dynamics \cite{Dydek2008, Gorinevsky2012}, and a noise-based signal may not be physically realizable due to constraints of physical actuators (Figure \ref{fig:inputs}).

Here, we propose a new approach for parameter estimation of \eqref{unmodeled-intro-x}-\eqref{unmodeled-intro-w} based on the theory of spectral lines, proposed in \cite{Boyd_1986,Boyd_1983}, in which a deterministic exogenous signal may be used to learn the parameters of the system, c.f. \cite{Anderson_1977,Anderson_1982,Narendra_1987,Willems2005}. 
These methods have been explored in depth in the field of adaptive control  \cite{Astrom_1995,Ioannou1996,Krstic_1995,Narendra2005,Sastry_1989}.
Our results prove useful for both learning the system \eqref{unmodeled-intro-x}-\eqref{unmodeled-intro-w}, as well as controlling the system, and we present theoretical results in the context of the Linear Quadratic Regular for the idealized system \eqref{lti-noisy-intro} which show competitive guarantees on Regret as considered in the literature \cite{Cohen2019, Mania2019}.

\paragraph{Related Work} The recent machine learning literature has addressed parameter estimation in linear dynamical systems in the context of control \cite{Dean_2018a, Dean_2018, Matni2019, Simchowitz_2019} and system identification \cite{Abbasi2011, Hardt_2016, Sarkar_2019, Sarkar_2019a, Farandobeh_2017,  Simchowitz_2018, Wagenmaker2020}. 
In the former, the goal is to show high-probability guarantees on the regret of the algorithm together with a Linear Quadratic Regulator in finite time \cite{Dean_2018a, Dean_2018, Matni2019, Simchowitz_2019}. 
In the latter,  papers such as \cite{Hardt_2016, Sarkar_2019,Farandobeh_2017,  Simchowitz_2018, Wagenmaker2020} have focused on system identification and provide high-probability guarantees of parameter estimates in finite time. 
Here, we show that the toolkit based on spectral lines can be used for both applications, and present our theorem statements in terms of both high probability guarantees for the systems \eqref{lti-noisy-intro} and \eqref{unmodeled-intro-x}-\eqref{unmodeled-intro-w} as well as in terms of Regret for the Linear Quadratic Regulator.

A closer examination of the results of two specific papers in the above list, \cite{Sarkar_2019, Wagenmaker2020}, is in order. Both of these works show estimation rates that decay as $\Tilde{O}(1/\sqrt{T})$. In Sarkar and Rakhlin \cite{Sarkar_2019}, the control input is taken to be a Gaussian input, 
\begin{equation} \label{white-noise-input}
u_k = s_k, \quad s_k \overset{i.i.d.}{\sim} \cN(0, \sigma_u^2 I) \,, \quad  \forall k \in \{1, \dots, T \}  \,.
\end{equation}
In contrast, Wagenmaker and Jamieson \cite{Wagenmaker2020} add a deterministic, periodic $\tilde{u}_k$ in addition to $s_k$. 
A numerical optimization procedure is used to actively determine $\tilde{u}_k$ in Wagenmaker and Jamieson \cite{Wagenmaker2020} and it is shown that it is advantageous to the passive approach in Sarkar and Rakhlin \cite{Sarkar_2019}. 
In contrast to these methods, our approach entirely removes the stochastic component, and proposes a spectral lines-based construction of the control input \cite{Boyd_1986, Boyd_1983}. 
The advantage of choosing the spectral lines-based input over the Gaussian input lies in robustness, avoidance of large inputs even at low probabilities, and active reduction in the parameter estimation error. 
The primary advantage of choosing the spectral line method of analysis over that suggested in Wagenmaker and Jamieson \cite{Wagenmaker2020} is its generalization to an arbitrary dynamic system, nonlinear, with unmodeled higher-order dynamics, as in \eqref{unmodeled-intro-x}-\eqref{unmodeled-intro-w}, that has a property of preserving spectral lines at its output. 
We leverage this property in providing estimation rates for this generalized class of dynamic systems. 
Algorithm \ref{alg:example} summarizes our approach.

\begin{algorithm}[H]
        % \caption{LQR Control with Gaussian Noise as Input}
        \caption{Example of LQR Control with Sub-Gaussian Spectral Lines}
        
        \label{alg:example}
        \begin{algorithmic}[1]
        % \STATE{ {\bfseries Require:} Stabilizing Controller $K^{(0)}$, failure probability $\delta \in (0, 1)$, Cost Matrices $Q, R$} \\
        \STATE{ {\bfseries Require:} Frequency Constraint Set $\mathcal{F}$, Stabilizing Controller $K^{(0)}$, failure probability $\delta \in (0, 1)$, Amplitude Constraint $M$, Cost Matrices $Q, R$} \\
        \FOR{$i=0,1,2,\ldots$}
        % \STATE{\textbf{Set} Epoch Time $T_i = C(A_*, B_*, Q, R, \delta) \times 2^i$, where $C(A_*, B_*, Q, R, \delta)$ is a constant function of underlying system parameters}
        \STATE{\textbf{Set} Epoch Time $T_i = C(A_*, B_*, Q, R, \delta) \times 2^i$, where $C(A_*, B_*, Q, R, \delta)$ is a constant function of underlying system parameters, \\
        Amplitude Constraint $\bar{M}^{(i)} = M T_i^{1/4}$}
        \STATE {\textbf{Choose} Distinct $f^{(i)}_1, \dots, f^{(i)}_{\lceil d / 2 \rceil} \in \mathcal{F},$\\ \qquad \qquad $M^{(i)}_1, \dots, M^{(i)}_{\lceil d / 2 \rceil}$, \\
        \qquad \qquad such that $\frac{1}{2} \bar{M}^{(i)}\leq M^{(i)}_j \leq \bar{M}^{(i)}$.}
        \FOR{$k = \sum_{\ell=0}^{i-1} T_\ell + 1, \dots, \sum_{\ell=0}^{i-1} T_\ell + T_i$}
        % \STATE {\textbf{Define} $u_{spec, k} = s_k, \quad s_k \overset{i.i.d.}{\sim} \cN(0, \sigma_u^2 I)$;}
        \STATE {\textbf{Define} $u_{spec, k} = \sum_{j = 1}^{\lceil d / 2 \rceil} M^{(i)}_j \cos(2 \pi f^{(i)}_j k)$;}
        \STATE {\textbf{Assign} $u_{k} = K^{(i)} x_k + u_{spec, k}$.}
        \STATE{ \textbf{Receive} $x_{k + 1} = A_*x_{k} + B_*u_{k} + w_k + \eta_{k}$}
        \ENDFOR
        \STATE{ \textbf{Estimate} \[(\hat{A}, \hat{B}) = \argmin_{A \in \R^{n\times n}, B \in \R^{n \times m}} \sum_{k = \sum_{\ell = 0}^{i-1} T_i + 1}^{\sum_{\ell=0}^{i-1} T_\ell + T_i} \norm{x_{k+1} - Ax_k - Bu_k}_2^2 \] }
        \STATE{ \textbf{Set} \[K^{(i+1)} = -(R + \hat{B}^\top \hat{P} \hat{B})^{-1} \hat{B}^\top \hat{P} \hat{A} \] for $\hat{P}$ which satisfies
        \[ \hat{P} = \hat{A}^\top \hat{P} \hat{A} - \hat{A}^\top \hat{P} \hat{B}  (R + \hat{B}^\top \hat{P} \hat{B})^{-1} \hat{B}^\top \hat{P} \hat{A} + Q\]}
        \ENDFOR
        \STATE{(Optional) In Line 4, select $M^{(i)}_1, \dots, M^{(i)}_{\lceil d / 2 \rceil}$, $f^{(i)}_1, \dots, f^{(i)}_{\lceil d / 2 \rceil}$ to optimize $\norm{\bar{\Phi}^{-1}}^{-1}$ as to minimize the bound in Theorem \ref{unmodeled-estimation}.} 
        \STATE{Note: In Lines 4-6, the specific form of $u_k$ can be altered as to include any signal which has sub-Gaussian spectral lines, such as pseudo-random binary sequences, as long as the amplitude of the signal is made to decay appropriately as in Line 3.} 
        \end{algorithmic}
\end{algorithm}

It should be noted that the computational burden associated with the input selection is non-existent, and that the frequencies need not be constrained; the amplitude selection in Line 4 is to constrain the excitation of the unmodeled dynamics, and therefore a better control design, and is otherwise not required for successful parameter estimation in \eqref{lti-noisy-intro}. 
This is in contrast to the active learning requirement imposed in \cite{Wagenmaker2020}. 
Other problem specific requirements that stem from magnitude and rate constraints on the exogenous input can be easily integrated into the algorithm if needed \cite{Morelli_2005}. 
It should also be noted that the proposed method has been analyzed at length in the system identification and adaptive control literature \cite{Goodwin_1984, Ljung_1987, Narendra2005} and has been highly successful in flight control applications \cite{Morelli_2005}.

\subsection{Summary of Results}
Our contributions in this work are as follows, where parentheses reference the specific section or result which details the contribution.

\begin{itemize}
\item We develop a new theory of spectral lines in a discrete-time and stochastic setting over a finite-time interval, bridging the gap between adaptive control and non-asymptotic estimation theory (Section \ref{s:spectral}).

\item We apply our results for the ideal system in \eqref{lti-noisy-intro}, and achieve competitive estimation rates with deterministic inputs as compared to the stochastic inputs in previous literature \cite{Abbasi2011, Sarkar_2019, Wagenmaker2020}. 
In particular, for the system \eqref{lti-noisy-intro}, we show that the user of the system has the ability to tune the estimation rate by selecting the spectral lines of the deterministic control input appropriately. 
Such flexibility is not provided in the previous literature that advocates the use of a stochastic input \cite{Abbasi2011, Dean_2018a, Dean_2018, Foster_2020, Matni2019, Sarkar_2019, Sarkar_2019a, Simchowitz_2019, Simchowitz_2018, Wagenmaker2020} (Section \ref{s:unmodeled}, Example \ref{ex:lti-estimation}).

\item In the case where $u_k$ is prescribed as a stochastic input for the system in \eqref{unmodeled-intro-x}-\eqref{unmodeled-intro-w}, we provide a lower bound which quantifies the extent to which the stochastic input can excite unmodeled dynamics ,resulting in inconsistent estimation.
Although the system in \eqref{unmodeled-intro-x}-\eqref{unmodeled-intro-w} is representative of true dynamics of physical systems, it has received little attention in recent literature. 
This indicates that existing results which hold for the idealized model \eqref{lti-noisy-intro} have limitations in practice, when unmodeled dynamics which have a deterministic relationship with the input are introduced (Section \ref{s:lower-bound}, Theorem \ref{lower-bound}).

\item In contrast, the approach proposed in this paper where $u_k$ is chosen appropriately using deterministic spectral lines for the system \eqref{unmodeled-intro-x}-\eqref{unmodeled-intro-w}, provides explicit estimation rates which depend on the frequency content $w_k$, and in practice can be minimized through proper choice of the spectrum of the exogenous signal.
As an illustrative example, we show that when $w_k$ is a high pass filter, parameter estimation error can be made arbitrarily small when the spectrum of $u_k$ is selected carefully, whereas the white noise input results in inconsistent estimation.
To our knowledge, we provide the first non-asymptotic estimation bound for a system of this form (Section \ref{s:unmodeled}, Theorem \ref{unmodeled-estimation}).

\item We show that, when the system dynamics are of the form \eqref{lti-noisy-intro} and the input $u_k$ is selected appropriately, in particular as in Algorithm \ref{alg:example}, the exploration results in $\tilde{O}(\sqrt{T})$ Regret with a Linear Quadratic Regulator based feedback controller, which is competitive in the literature (Section \ref{s:lqr-regret}, Theorem \ref{thm:regret}).

\item We provide experimental results which indicate the efficacy of using a spectral-lines based exogenous input compared to an input based on white noise, in terms of both parameter estimation error and closed-loop control design using the Linear Quadratic Regulator (Section \ref{s:experiments}).

\end{itemize}

Throughout this work, we will make the following assumption on the systems \eqref{lti-noisy-intro} and \eqref{unmodeled-intro-x}-\eqref{unmodeled-intro-w}.
\begin{assumption} \label{assumption-stability}
In \eqref{lti-noisy-intro} and \eqref{unmodeled-intro-x}-\eqref{unmodeled-intro-w}, the matrix $A_*$ is assumed to be Schur-stable.
\end{assumption}
To ensure that the signals of interest are well behaved, we make this standard assumption, commonly satisfied by physical systems. As articulated in \cite{Ljung_1987}, closed-loop system identification based methods can be used to relax this requirement. It is also worth noting that the developments in \cite{Sarkar_2019} suggest that Assumption \ref{assumption-stability} is not required for the statistical analysis of parameters in the system \eqref{lti-noisy-intro}. As our goal is to focus on methods that are of interest in physical dynamic systems, we retain this assumption; relaxations of this assumption using methods in \cite{Ljung_1987, Sarkar_2019} are straight forward.

% A few specific observations are made in comparison with bounds on parameter estimation derived here with those considered in \cite{Sarkar_2019, Wagenmaker2020} are in order. Both of these works show estimation rates that decay as $\Tilde{O}(1/\sqrt{T})$. In \cite{Sarkar_2019}, the control input is taken to be $u_k \overset{i.i.d.}{\sim} \cN(0, I)$, whereas \cite{Wagenmaker2020} takes an adaptive approach based on periodic signals to the selection of $u_k$ in addition to a Gaussian input, and hence provides an analysis of an active learning method as opposed to an analysis of fixed $u_k$ a priori. Similar to \cite{Wagenmaker2020}, this work considers the use of periodic signals to generate excitation; however, our approach differs in that we provide an explicit generalization of the notion of spectral lines %which have been widely studied in adaptive control%
% \cite{Boyd_1986, Boyd_1983} and, more importantly, we provide an analysis for any fixed sequence $u_k$ as opposed to an adaptively selected exogenous input. We also do not include Gaussian noise in the input, thus maintaining the physical feasibility of implementation of the input in real systems and leading to better robustness.

We begin with a discussion of preliminaries in Section \ref{s:Preliminaries} to introduce notions from adaptive control, including the theory of spectral lines and how they elicit excitation, as well as notions from non-asymptotic statistics, which have been used widely in recent literature \cite{Abbasi2011, Dean_2018, Sarkar_2019, Sarkar_2019a}. 
In Section \ref{s:spectral}, we then provide our new definitions regarding spectral lines in discrete-time, stochastic settings. 
Section \ref{s:unmodeled} the key estimation theorem leveraging the new notions of deterministic spectral lines for the system in \eqref{unmodeled-intro-x}-\eqref{unmodeled-intro-w}, and provide the corresponding lower bound for when the input is chosen according to white noise.
Section \ref{s:lqr-regret} then applies our results to the context of the Linear Quadratic Regulator, in which we show competitive regret bounds using the spectral-lines based approach.
Section \ref{s:experiments} concludes with simulated results indicating that selecting a control input based on spectral lines outperforms a white noise-based control input in the presence of unmodeled dynamics. 
\paragraph{Notation} Let $\R$ denote the set of real numbers $\R^+$ denote the set of non-negative real numbers, and $\R^n$ denote the set of real-valued vectors of length $n$. We let $\mathcal{S}^{d-1}$ denote the unit sphere in $\R^d$. The functions $\Re(\cdot)$ and $\Im(\cdot)$ represent the real and imaginary parts of their inputs, respectively. Let $\Omega_T = \{0, 1/T, \dots, (T-1)/T \}$ be the finite set of discrete frequencies for a sequence of length $T$. Given a real-valued, finite sequence $\{y_k\}_{k = 0}^{T-1}$, we denote the Discrete Fourier Transform of the sequence using a bolded letter as $\y (e^{j\omega}) = \sum_{k = 0}^{T-1} y_k e^{-j2 \pi k\omega}$, where $\omega \in \Omega_T$. Let $\norm{\cdot}$ represent the Euclidean norm when its input is a vector, and the operator norm when its input is a matrix. Further, for a matrix $B \succeq 0$, we define $\norm{A}_B = \norm{B^{1/2} A}$.

\section{Preliminaries}
\label{s:Preliminaries}
In this section, we introduce the main technical tools employed in this paper. In Section \ref{ss:adaptive_preliminaries} we discuss notions of persistent excitation and spectral lines for parameter convergence, as is common in the field of adaptive control theory. We then proceed to a discussion of non-asymptotic estimation tools from the recent literature in Section \ref{ss:stats_preliminaries}.

\subsection{Parameter Convergence in Discrete Time Linear Regression}
\label{ss:adaptive_preliminaries}

To motivate a later discussion of spectral lines and excitation conditions, we first consider the problem of linear regression with time-varying regressors in discrete time, which has a large body of work in both the system identification and the adaptive control literature \cite{Anderson_1982, Bai1984, Goodwin_1984,Ljung_1987,Narendra2005}. In the Appendix, we provide a brief overview of well-studied tools in the continuous time setting, which we hope will inform future research on the subject in the context of machine learning. 

The discrete time linear regression setting with a time-varying regressor can be briefly stated as the estimation of the parameter $\theta_*$ in the regression relation
\begin{equation}
y_k=\theta_*^{\top}\phi_{k} \,,
\label{daa1}
\end{equation}
where $y_k \in \R$ is an observed outcome, $\phi_k \in \R^N$ is the time-varying regressor, and $\theta_* \in \R^N$ is unknown. Such a time-varying regression setting has been common in the adaptive control literature, and encapsulates a variety of well-studied settings such as the ARMA model \cite{Goodwin_1984, Ljung_1987, Whittle1951}. Given that $\theta_*$ is unknown, we formulate an estimator $\hat{y}_k = \theta_k^\top \phi_{k}$, where $\theta_k\in\mathbb{R}^N$ is an adjusted parameter and $\hat{y}_k\in\mathbb{R}$ is the predicted output. For example, in Anderson and Johnson \cite{Anderson_1982}, the underlying estimator that is used is
\begin{equation}\label{anderson-est}
\theta_{k+1}=\theta_k -\gamma \frac{1}{\cN_k}(\theta_k^\top \phi_{k}-y_k) \,,
\end{equation}
where $\cN_k$ is a suitably chosen normalization which guarantees that $\theta_k$ remains bounded for all initial conditions $\theta_{t_0}$, and $0<\gamma < 2$ \cite[Chapter 3]{Goodwin_1984}. %Here, it is worth noting the online learning aspects of the parameter estimation algorithm are underscored as the regressor-output data pairs $(\phi_k,y_k)$ are streaming and the parameter estimate $\theta_k$ is updated at each discrete time step $k$.
The output prediction error is of the form
\begin{equation}\label{discrete-error}
    e_k = \hat{y}_k-y_k =\tilde{\theta}_k^\top \phi_{k} \,,
\end{equation}
where $\tilde{\theta}_k = \theta_k - \theta_*$ is the parameter estimation error. 
% \begin{example}\label{ex:arma}
% Consider the setting where $y_k$ is the output of an ARMA \cite{Whittle1951} model whose regressor $\phi_{k}$ is given by the past inputs and outputs, and is of the form
% \begin{equation}
% \label{dreg}
% \phi_k=[y_k, y_{k-1},\ldots, y_{k-n+1}, u_k, u_{k-1},\ldots, u_{k-m}]^\top \,,
% \end{equation}
% where $n$ represents the number of previous states influencing the current state and $m$ represents the number of previous control inputs which influence the current state.
% \end{example}
% The preceding example of time-varying regression has been widely studied in the adaptive control literature with a breadth of applications \cite{Anderson_1982, Goodwin_1984, Ljung_1987}, and forms the basis of many methods recently considered in machine learning for system identification \cite{Abbasi2011, Dean_2018a, Dean_2018, Foster_2020, Matni2019, Sarkar_2019, Sarkar_2019a, Simchowitz_2019, Simchowitz_2018, Wagenmaker2020}.
A primary goal of time-varying regression is to ensure that, as $k \rightarrow \infty$, the parameter estimation error $\tilde{\theta} \rightarrow 0$ as well. A secondary goal would be to at least ensure that, as $k \rightarrow \infty$, the output prediction error $e_k \rightarrow 0$. While the primary goal ensures the latter, the converse implication is not necessarily true. We may, however, begin to bridge the gap between the two goals by introducing the following necessary and sufficient condition for these classes of problems.
\begin{definition}\cite{Bai1984} \label{discrete-pe}
A regressor $\{ \phi_k \}_{k=1}^\infty$ is said to be \textit{persistently exciting} if there exist strictly positive constants $\rho_1<\rho_2$, and integers $k_0, S \geq 1$ such that
\begin{equation}
\rho_2I \succeq \sum_{k=j}^{j+S} \phi_k \phi_k^\top \succeq \rho_1I  \,,
\end{equation}
for all $j \geq k_0$.
\end{definition}
In particular, the definition above allows for the following result showing the importance of persistent excitation.
\begin{theorem}\cite[Chapter 3.4]{Goodwin_1984}
Consider the estimator \eqref{anderson-est}. If the regressor $\{\phi_k \}_{k = 1}^\infty$ satisfies the condition in Definition \ref{discrete-pe}, then for any $\epsilon > 0$, there exists a $T = O(\log \epsilon)$ such that
\begin{equation}\label{discrete-conv}
\norm{\theta_k-\theta^*} \leq \epsilon,\qquad \forall k\geq T.
\end{equation}
\end{theorem}
There is an intimate relationship between whether or not an input satisfies the condition of persistent excitation and its spectral content, quantified using \emph{Spectral Lines}:
\begin{definition}\cite{Bai1984} \label{discrete-spectral-line}
Consider $\{ \phi_k \}_{k=0}^\infty$ and a value $\nu$. Then, $\{ \phi_k \}_{k=0}^\infty$ has a spectral line at $\nu \in [-\pi, \pi]$ with amplitude $\bar{\phi}(j \nu)$ if 
\[ \lim_{M \rightarrow \infty} \frac{1}{M} \sum_{k = k_0 + 1}^{k_0 + M} \phi_k e^{-j \nu k} = \bar{\phi} (j \nu) \,,\]
uniformly in $k_0$.
\end{definition}
In particular, it is known that if the spectral content of a regressor has sufficiently many spectral lines with linearly independent amplitudes, then the persistent excitation condition in Definition \ref{discrete-pe} is immediate \cite{Bai1984}. That is, if the spectral content of the regressor spans sufficiently many frequencies, then we are able to show that the parameter estimation error $\tilde{\theta}$ tends to 0 when $\theta_k$ is updated according to \eqref{anderson-est} \cite{Anderson_1982, Bai1984}. 

Recent advances in adaptive control have begun to consider the setting in which there exists only a finite amount of data, where $k \in \{1, \dots, T\}$  \cite{Cho_2018,Gaudio2019Parameter,Jha2019}, which leads to \emph{Finite Excitation}.
\begin{definition}[Finite Excitation]\label{discrete-fe}
A regressor $\{ \phi_k \}_{k=i}^{i + S}$ is said to be \textit{finitely exciting} from $i$ to $i + S$ if there exist strictly positive constants $\rho_1<\rho_2$ such that
\begin{equation}\label{dpe}
\rho_2I \succeq \sum_{k=i}^{i+S} \phi_k \phi_k^\top \succeq \rho_1I  \,.
\end{equation}
\end{definition}
As we will see in Section \ref{s:spectral}, the notion of finite excitation helps to bridge the asymptotic results of adaptive control with the non-asymptotic setting which has been considered in recent literature in machine learning.

\subsection{Non-Asymptotic Estimation Bounds}
\label{ss:stats_preliminaries}

In order to accommodate settings with finite data and unobservable external disturbances, we leverage tools from recent advances in non-asymptotic statistics \cite{Vershynin2019}  and propose non-asymptotic estimation bounds in this section. First, we note the definition of a sub-Gaussian random variable and a sub-Gaussian random vector.
\begin{definition}\cite{Vershynin2019}
A random variable $X$ is sub-Gaussian with variance proxy $\sigma^2$ if there exists $\sigma^2 > 0$ such that
\[ \E[e^{\lambda X}] \leq e^{\frac{\lambda^2 \sigma^2 }{2}}, \quad \forall \lambda > 0 \,.\]
The relationship above is denoted $X \sim \subG(\sigma^2)$.
\end{definition}
\begin{definition}\cite{Pisier2016}
A complex valued vector random variable $X$ which takes values  in $\R^{d}$ is said to be a sub-Gaussian (Exponential) with parameter $R^2$ if, for all $z \in \mathcal{S}^{d-1}$, the random variables $\Re (z^\top X)$ and $\Im (z^\top X)$ are sub-Gaussian with parameter $R^2$. If a random vector satisfies the above definition, then we say $X \sim \subG(R^2)$.
\end{definition}
These definitions then lead to a self-normalized martingale bound, shown by Abbasi-Yadkori et al. \cite{Abbasi2011}, which we will leverage for system identification.
% \begin{theorem}[Theorem 1 in \cite{Abbasi2011}]\label{th:MartingaleTheorem}
% Let $\{\cF_k \}_{k=1}^\infty$ be a filtration. Let $\{\eta_k\}_{k=1}^\infty$ be a sequence of real-valued random variables such that $\eta_k$ is $\cF_{k+1}$-measurable and conditionally $R^2$ sub-Gaussian given $\cF_{k}$, i.e.
% \[ \E[e^{\lambda\eta_k} | \cF_{k}] \leq e^{\frac{\lambda^2 R^2}{2}}, \qquad \forall~\lambda > 0 \,. \]
% Let $\{ x_k \}_{k=1}^\infty$ be a sequence of random vectors such that $x_k \in \R^d$ is $\cF_k$ measurable, and let $V \in \R^{d \times d}$ be an arbitrary positive definite deterministic matrix, and define
% \[ \bar{V}_T = V + \sum_{k = 1}^T x_k x_k^\top, \qquad S_T = \sum_{k = 1}^T x_k \eta_{k}  \,. \]
% Then, for any $\delta > 0$, with probability at least $1 - \delta$, for all $T > 0$, 
% \[ \norm{S_T}_{\bar{V}_T}^2 \leq 2R^2 \log \left( \frac{\det(\bar{V}_T)^{1/2} \det(V)^{-1/2}}{\delta} \right).\]
% \end{theorem}
Theorem 1 of Abbasi-Yadkori et al. \cite{Abbasi2011}, which is restated in the Appendix, has been crucial to the analysis of linear system identification in the recent machine learning literature \cite{Dean_2018, Sarkar_2019, Simchowitz_2018, Wagenmaker2020}. In particular, the theorem allows for the analysis of the martingale terms that arise in the context of linear dynamical systems. Much like Proposition 3 of Sarkar and Rakhlin \cite{Sarkar_2019}, we are able to extend the result of Abbasi-Yadkori et al. \cite{Abbasi2011} to the setting of \eqref{unmodeled-intro-x}-\eqref{unmodeled-intro-w} in the following proposition, the proof of which can be found in the Appendix.
\begin{proposition}[Self-Normalized Martingale Bound] \label{self-norm-body}
Consider the system \eqref{unmodeled-intro-x}-\eqref{unmodeled-intro-w} where $w_k, \eta_k, x_k \in \R^n$, $\eta_k \sim \subG(\sigma^2)$, and $u_k \in \R^m$ is deterministic. For an arbitrary deterministic matrix $V \succ 0$, define
\[ \phi_k = \begin{bmatrix}
x_k \\ u_k
\end{bmatrix}, \quad \bar{Y}_T = V + \sum_{k=1}^T \phi_k \phi_k^\top, \quad S_T = \sum_{k = 1}^T \phi_k \eta_k^\top \,. \]
Then, for any $0< \delta < 1$, with probability at least $1 - \delta$,
\begin{align}
    &\norm{(\bar{Y}_T)^{-1/2} S_T}  \\&\leq\sigma \sqrt{8d \log \left( \frac{5 \det(\bar{Y}_T)^{1/(2d)} \det(V)^{-1/(2d)}}{\delta^{1/d}} \right)} \,,
\end{align} 
where $d = n + m$.
\end{proposition}
Since the system \eqref{lti-noisy-intro} is a special case of the system \eqref{unmodeled-intro-x}-\eqref{unmodeled-intro-w} where $w_k = 0$ for all $k$ in \eqref{unmodeled-intro-w}, we see the above proposition applies to both systems.
With the toolkit for the theory of spectral lines from adaptive systems in Section \ref{ss:adaptive_preliminaries} and recent results from non-asymptotic statistic in Section \ref{ss:stats_preliminaries}, we are able to provide estimation bounds for the system \eqref{unmodeled-intro-x}-\eqref{unmodeled-intro-w}.
As a special case, this allows us to achieve competitive regret bounds for the Linear Quadratic Regulator, for which we provide some background below.

\subsection{Regret Bounds for the Linear Quadratic Regulator}
\label{ss:regret_preliminaries}

The Linear Quadratic Regulator (LQR) problem has been studied at length in the control literature \cite{Ljung_1987}, and has received much recent attention in the machine learning literature \cite{Abbasi2011, Cohen2019, Dean_2018, Mania2019}.
Formally, the problem is stated as follows for positive definite matrices $Q$ and $R$:
\begin{equation} \label{eq:lqr}
    \min_{u_k} \lim_{T \rightarrow \infty} \mathbf{E}  \left[  \frac{1}{T}\sum_{k = 0}^T x_k^\top Q x_k + u_k^\top R u_k \right] \qquad \text{s.t. dynamics \eqref{lti-noisy-intro}} \,.
\end{equation}
We denote this minimizing quantity as $J_*$, and note that it can be achieved when $A_*$ and $B_*$ are known through the solution of the Ricatti equation to determine an optimal controller $K_*$, such that the optimal strategy is to assign $u_k = K_* x_k$.
When $A_*$ and $B_*$ are unknown, the performance of an algorithm is measured with regret,
\begin{equation}\label{eq:regret}
    \textrm{Regret}(T) = \sum_{k = 0}^T (x_k^\top Q x_k + u_k^\top R u_k - J_*) \,.
\end{equation}
Recent work has shown that the certainty equivalence approach, which solves the Ricatti equation using estimates $\hat{A}$ and $\hat{B}$, achieves small regret when $\hat{A}$ and $\hat{B}$ are sufficiently close to their true values \cite{Mania2019}.

Specifically, the authors make the following assumption about the system \eqref{lti-noisy-intro}.
\begin{assumption}[$(\ell, \nu)$-stability] \label{assumption:strong-stability}
The matrices $A_*$ and $B_*$ satisfy
\[ \sigma_{\textrm{min}} \left( \begin{bmatrix} B_* & A_* B_* & \dots & A_*^{\ell - 1} B_*\end{bmatrix} \right) \geq \nu \,, \]
for some integer $\ell$ and real number $\nu > 0$.
\end{assumption}
This assumption is slightly stronger than the assumption of controllability.

With this assumption, the authors are able to show the following claim:
\begin{proposition} \label{prop:estimation-to-regret}
Suppose $A_*, B_*$ satisfy Assumption \ref{assumption:strong-stability} for an integer $\ell$ and constant $\nu > 0$, that the cost matrices $Q$ and $R$ are positive semi-definite, and that $\max\{\norm{\hat{A} - A_*}, \norm{\hat{B} - B_*}\} \leq \epsilon$ for some $0 < \epsilon < \bar{\epsilon}(A_*, B_*, Q, R)$ where $\bar{\epsilon}(A_*, B_*, Q, R)$ is a function which depends only on underlying system parameters.
Let $\hat{J}$ represent the cost 
\[ \lim_{T \rightarrow \infty} \mathbf{E}  \left[  \frac{1}{T}\sum_{k = 0}^T x_k^\top Q x_k + u_k^\top R u_k \right] \qquad \text{s.t. dynamics \eqref{lti-noisy-intro}} \,,\]
when $u_k = \hat{K} x_k$, where $\hat{K}$ is the solution of the Ricatti equation with inputs $(\hat{A}, \hat{B})$.
Then, for sufficiently small $\epsilon$ as determined by $(A_*, B_*, Q, R)$, the following holds:
\begin{equation}\label{eq:regret-result}
    \hat{J} - J_* \leq \tau(A_*, B_*, Q, R) \epsilon^2 \,,
\end{equation}
where $\tau(A_*, B_*, Q, R)$ is a constant that depends only on system parameters.
\end{proposition}
This result quantifies the extent to which accurate estimation results in risk minimization for the LQR problem.
In Section \ref{s:lqr-regret}, we explicitly show how efficient estimation rates combined with Proposition \ref{prop:estimation-to-regret} above can result in $\tilde{O}(\sqrt{T})$ regret in control of the linear quadratic regulator.
First, we establish the efficient estimation bounds by introducing the theoretical toolkit of sub-Gaussian spectral lines.

\section{A New Theory for Spectral Lines of Discrete Stochastic Signals}
\label{s:spectral}
The focus of this paper is on  the notion of spectral lines over a finite time-interval. Towards this end, we introduce Definition \ref{discrete-stochastic-spectral} that defines a spectral line based on finite sample of data in a stochastic setting.
\begin{definition}[Sub-Gaussian Spectral Line]
\label{discrete-stochastic-spectral}
A stochastic sequence $\{ \phi_k \}_{k \geq 0}$ is said to have a sub-Gaussian spectral line from $i$ to $i + S$ at a frequency $\omega_0$ of amplitude $\bar{\phi}(\omega_0)$ and radius $R$ if
\begin{equation}
\label{eq:sg-spec-line}
 \frac{1}{S + 1} \sum_{k=i}^{i+S} \phi_k e^{-j 2\pi \omega_0 k } - \bar{\phi}(\omega_0) \sim \subG(R^2 / (S+1)) \,.
\end{equation}
\end{definition}
Definition \ref{discrete-stochastic-spectral} determines the frequency content of a stochastic signal by decoupling the part of the signal which is deterministic from the part of the signal directly affected by the stochastic process noise. Hence, we may decouple our analysis and apply tools from adaptive control to the deterministic part of the signal while using tools from non-asymptotic statistics to place bounds on the stochastic part of the signal with particular focus on bounding aberrant behavior with high probability. Examples of sequences which have sub-Gaussian spectral lines include deterministic discrete sinusoids, such as those used in Algorithm \ref{alg:example}, which will have a sub-Gaussian spectral line of radius 0 due to their deterministic nature. One can generate stochastic signals with sub-Gaussian spectral lines as well, for example through the summation of a discrete sinusoid with a Gaussian signal as used in Wagenmaker and Jamieson \cite{Wagenmaker2020}.
\begin{remark}
Definition \ref{discrete-spectral-line} pertains to the notion of a spectral line, and requires the underlying signal $\{ \phi_k \}_{k \geq 0}$ to be specified for all $k \in \mathbb{N}$. In contrast, Definition \ref{discrete-stochastic-spectral} introduces the notion of a sub-Gaussian spectral line where it suffices for $\{ \phi_k \}_{k \geq 0}$  to be specified over a finite time-interval.  This is a stronger condition on $\{ \phi_k \}_{k \geq 0}$, as we require the signal to have the appropriate behavior over a finite duration.
\end{remark}
We leverage Definition \ref{discrete-stochastic-spectral} in the following lemma, which relates the input and output of a system to one another in terms of their spectral content. The following new lemma is a discrete time, stochastic analogue to that provided by Lemma 3.3 of Boyd and Sastry \cite{Boyd_1983}.
\begin{lemma}
\label{spectral-transfer}
Consider $\{u_k\}_{k = 0}^{T-1}, \{ y_k \}_{k=0}^{T-1} $ as the input and output, respectively, of a discrete-time, stable linear time-invariant system with arbitrary initial conditions and an unobserved external disturbance. Specifically, suppose $\y(e^{j\omega})$ and $\u(e^{j\omega})$ are related as
\begin{equation}
\y(e^{j\omega}) = H(e^{j\omega}) \u(e^{j\omega}) + \boldeta (e^{j\omega}),
\end{equation}
where $H(e^{j\omega})$ is a deterministic transfer function and $\boldeta (e^{j\omega})\sim \subG(f(e^{j\omega})^2)$. If $\{u_k\}_{k = 0}^{T-1}$ has a sub-Gaussian spectral line at frequency $\omega_0 \in \Omega_T$ from $0$ to $T-1$ with amplitude $\bar{u} (\omega_0)$ and radius $R$, then $\{y_k\}_{k=0}^{T-1}$ has a sub-Gaussian spectral line from $0$ to $T-1$ with amplitude
\begin{equation}
\bar{y}(\omega_0) = H(e^{j \omega_0}) \bar{u} (\omega_0)
\end{equation}
and radius $||H(e^{j\omega_0})|| R + f(e^{j\omega_0})$.
\end{lemma}
A proof of Lemma \ref{spectral-transfer} can be found in the Appendix. 
%It is worth noting that, while Definition \ref{discrete-stochastic-spectral} can potentially be defined for any frequency, the use of the DFT in the transfer function restricts the application of the lemma to frequencies in the finite set $\Omega_T$. 
Lemma \ref{spectral-transfer} shows that the spectral content of the exogenous signal $\{u_k\}_{k=0}^{T-1}$ affects the spectral content of $\{y_k\}_{k=0}^{T-1}$ in a natural way, with the amplitude of the spectral line determined by the bandwidth of the system, and the radius being affected by external noise and stochasticity of $\{u_k\}_{k=0}^{T-1}$ itself.

In order to relate the spectral content of a signal to the necessary and sufficient condition of persistent excitation provided by Boyd and Sastry \cite{Boyd_1983}, we provide a new discrete time stochastic analogue to Lemma 3.4 of Boyd and Sastry \cite{Boyd_1983}. This proposition relates the number of spectral lines with linearly dependent amplitudes of a sequence to whether or not such a sequence is finitely exciting. In order to show this claim, we must first define the expected information matrix of a sequence of stochastic vectors.
\begin{definition}[Expected Information Matrix] \label{big-phi}
Let $\phi_k \in \R^n$ be a sequence of stochastic vectors. If $\phi_k$ has $n$ sub-Gaussian spectral lines at frequencies $(\omega_1, \dots, \omega_n) := \Omega$ from $i$ to $i + S$ with amplitudes $\{ \bar{\phi}(\omega_1), \dots, \bar{\phi} (\omega_n) \}$, then the expected information matrix $\bar{\Phi}$ is defined
\[  \bar{\Phi} = \begin{bmatrix}
\vert & \dots & \vert \\
\bar{\phi}(\omega_1) & \dots &\bar{\phi} (\omega_n) \\
\vert & \dots & \vert \\
\end{bmatrix} \,. \]
\end{definition}
The expected information matrix $\bar{\Phi}$ in Definition \ref{big-phi} represents the core idea in any system identification problem. In the deterministic setting it is well known that a full rank and well conditioned $\bar{\Phi}$ results in fast estimation of unknown parameters. With the information matrix defined, we can show a clear relationship between the spectral content of a signal and whether it is finitely exciting.
\begin{proposition}
\label{spectral-to-PE}
Let $\phi_k \in \R^n$ be a sequence of stochastic vectors. If $\phi_k$ has $n$ spectral lines at frequencies $\omega_1, \dots, \omega_n$ from $i$ to $i + S$ with amplitudes $\{ \bar{\phi}(\omega_1), \dots, \bar{\phi} (\omega_n) \}$ which are linearly independent in $\C^n$, and maximum radius $R$ as defined in Definition \ref{discrete-stochastic-spectral}, then $\phi_k$ is finitely exciting from $i$ to $i + S$ with probability at least $1 - e^{-\frac{c \norm{\bar{\Phi}^{-1}}^{-1} (S + 1)}{2nR^2} + 2n \log 9},$ where $c$ is an absolute constant. Specifically, with probability at least $1 - \delta$, $\phi_k$ will satisfy
\[ \sum_{k = i}^{i+S} \phi_k \phi_k^\top \succeq \frac{1}{2n} \norm{\bar{\Phi}^{-1}}^{-2}I \,, \]
so long as $S \gtrsim R^2 (\log (1/\delta) + \norm{\bar{\Phi}^{-1}})$.
\end{proposition}
\begin{remark}
In Section \ref{s:unmodeled}, a deterministic choice of $u_k$ will be shown to ensure that the expected information matrix $\bar{\Phi}$ is well-conditioned. In contrast, the recent literature (e.g. \cite{Abbasi2011, Dean_2018a, Dean_2018, Matni2019, Sarkar_2019, Wagenmaker2020}) selects the control input as in \eqref{white-noise-input}, as opposed to a deterministic spectral-line based exogenous signal. This choice of such a zero-mean exogenous signal sets $\bar{\Phi} = 0$. The analysis in this work quantifies the impact of $u_k$ on $\bar{\Phi}$, which allows for tunable estimation rates.
\end{remark}

% \begin{remark}
%     The expected information matrix $\bar{\Phi}$ in Definition \ref{big-phi} represents the core idea in any system identification problem. It is well known in the deterministic setting that a full rank and numerically well conditioned $\bar{\Phi}$ results in fast estimation of unknown parameters in system identification. In the recent literature in machine learning, (e.g. \cite{Abbasi2011, Dean_2018a, Dean_2018, Foster_2020, Matni2019, Sarkar_2019, Sarkar_2019a, Simchowitz_2019, Simchowitz_2018, Wagenmaker2020}), $\bar{\Phi} = 0$, and any well-conditioning of the information matrix results from high-probability guarantees on the behavior of the random inputs $u_k$. In this paper we depart from this approach by using a deterministic choice for the input $u_k$ to center the distribution of the information matrix around a full rank $\bar{\Phi}$.
% \end{remark}
The proof of Proposition \ref{spectral-to-PE} may be found in the Appendix. Proposition \ref{spectral-to-PE} allows us to bridge the gap between the discrete time and deterministic setting of Bai and Sastry \cite{Bai1984} which has been considered widely in the adaptive control literature \cite{Goodwin_1984,Ljung_1987,Narendra2005}, with the more recently considered stochastic settings of the machine learning literature in which the exogenous signal is stochastic \cite{Abbasi2011, Dean_2018a, Dean_2018, Matni2019, Sarkar_2019, Tsiamis_2019}. In particular, Proposition \ref{spectral-to-PE} can be viewed as the non-asymptotic, stochastic equivalent condition of persistent excitation, and leverages the notion of spectral lines as in \cite{Boyd_1983}. The latter in turn helps us establish tunable estimation rates for parameters in both linear time-invariant systems and general dynamic systems, as outlined in Section \ref{s:unmodeled}.

% \section{Tunable Estimation Rates in Linear Time Invariant Systems}
% \label{s:lti}
% \input{body/lti-estimation.tex}

\section{Estimation Bounds in the Presence of Unmodeled Dynamics}
\label{s:unmodeled}
In this section, we apply the definitions above to consider the system \eqref{unmodeled-intro-x}-\eqref{unmodeled-intro-w}, where the influence of the spectrum of $u_k$ on $x_k$ may be complex due to the presence of unmodeled dynamics $w_k$. By exploiting the structure of \eqref{unmodeled-intro-x}-\eqref{unmodeled-intro-w}, we may rewrite the system in the frequency domain as
\begin{equation} \label{unmodeled-freq-domain}
\begin{aligned}
    \x(e^{j \omega}) &= (e^{j \omega}I - A_*)^{-1}B_* \u(e^{j \omega}) + \\&\quad (e^{j \omega}I - A_*)^{-1} \bigg(\w(e^{j \omega}) + \boldeta(e^{j \omega}) \bigg)\,.
\end{aligned}
\end{equation}
In order for the above statement to hold, we make the following technical assumption.
\begin{assumption} \label{wk-assumption}
In \eqref{unmodeled-intro-x}-\eqref{unmodeled-intro-w}, $f$ is Lipschitz \cite{Rudin1987}. %, i.e. for any deterministic and bounded choice of $\{u_k\}_{k=0}^{T-1}$, the sequence $\{w_k\}_{k=0}^{T-1}$ is deterministic and bounded.
\end{assumption}
In particular, Assumption \ref{wk-assumption} implies that, as long as $\{u_k\}_{k=0}^{T-1}$ is Fourier integrable in the discrete setting, then so is $\{w_k\}_{k=0}^{T-1}$ \cite{Oppenheim1999}. Under this condition, when $u_k$ is deterministic we know that $w_k$ will have sub-Gaussian spectral lines with radius 0, as it will also be deterministic and will have a well-defined discrete Fourier Transform.
We then see from \eqref{unmodeled-freq-domain} and Lemma \ref{spectral-transfer}, for $\phi_k = [x_k^\top u_k^\top]^\top$, the sequence $\{\phi_k\}_{k=0}^{T-1}$ has a spectral line at frequency $\omega_0$ from $0$ to $T - 1$ with amplitude
\begin{align*}
    \bar{\phi}(\omega_0) = \begin{bmatrix}(e^{j \omega}I - A_*)^{-1} B_* \\ I_m \end{bmatrix} \bar{u}(\omega_0) ~~+ \\ \begin{bmatrix}
(e^{j\omega}I - A_*)^{-1}  \\
0
\end{bmatrix} \bar{w}(\omega_0) \,, 
\end{align*} 
and radius $\sigma^2$. Hence, in the case where $u_k$ is deterministic, the information matrix for the system is clearly defined leading to the following result.
\begin{theorem} \label{unmodeled-estimation}
Consider the system \eqref{unmodeled-intro-x}-\eqref{unmodeled-intro-w} under Assumption \ref{wk-assumption}. Suppose the control input $u_k$ is selected deterministically such that $\phi_k = [x_k^\top u_k^\top]^\top$ has $n + m$ spectral lines with linearly independent amplitudes and finite radii. Then, the least-squares estimates $(\hat{A}, \hat{B})$ defined as
\begin{equation} \label{least-squares}
    (\hat{A}, \hat{B}) = \argmin_{A \in \R^{n\times n}, B \in \R^{n \times m}} \sum_{k = 0}^{T-1} \norm{x_{k+1} - Ax_k - Bu_k}_2^2 \,,
\end{equation} 
will satisfy, with probability at least $1 - \delta$, 
\begin{align}
     &\max \left\lbrace \norm{\hat{A} - A_*}, \norm{ \hat{B} - B_*}\right\rbrace \leq \label{system-id-thm}\\ 
     &\qquad\tilde{O} \left( \sqrt{\frac{1}{T \norm{\bar{\Phi}^{-1}}^{-2}}} + \frac{1}{T \norm{\bar{\Phi}^{-1}}^{-2}} \left(\sum_{k = 0}^{T-1} \norm{\bar{\phi} (e^{jk/T}) \mathbf{w} (e^{jk/T})^\top } + \frac{1}{\sqrt{T}}  \norm{\mathbf{w} (e^{jk/T})}\right) \right) \nonumber \,,
\end{align}
where the $\tilde{O}$ notation hides terms associated with dimensions $n$ and $m$, as well as terms logarithmic in $T$ and $\frac{1}{\delta}$.
\end{theorem}
A proof of Theorem \ref{unmodeled-estimation} can be found in the Appendix. %In order to show the utility of Theorem \ref{unmodeled-estimation}, we must consider the form of the quantity $m_T$, which the practitioner would wish to minimize. This quantity may not necessarily be small for all choices of $f(\cdot)$ or arbitrary selections of $u_k$. However, $m_T$ can be shown to be of small order in several cases, as will be shown in several examples. In cases in which $m_T$ may not be allowed to decay appropriately, it is worth noting that a system of the from \eqref{unmodeled-intro-x}-\eqref{unmodeled-intro-w} may not be the appropriate model for estimation.
\begin{example}[Linear Time Invariant Systems] \label{ex:lti-estimation}
In the special case in which $w_k$ is identically equal to 0, the system \eqref{unmodeled-intro-x}-\eqref{unmodeled-intro-w} is equivalent to the system \eqref{lti-noisy-intro}, which has been widely considered in the recent machine learning literature \cite{Abbasi2011, Dean_2018a, Dean_2018, Matni2019, Sarkar_2019, Sarkar_2019a, Wagenmaker2020}. The estimation rate can be written
\[ 
    \max \left\lbrace ||\hat{A} - A_*||, || \hat{B} - B_*|| \right\rbrace \leq \tilde{O} \left( \sqrt{\frac{1}{T \norm{\bar{\Phi}^{-1}}^{-2}}} \right)\,.
\]
\end{example}
In particular, we recover the same asymptotic rate of estimation of \cite{Sarkar_2019, Wagenmaker2020} by using a simple proof technique foundational to the analysis of spectral lines \cite{Boyd_1983}.  It is worth noting, that unlike previous work on the estimation of linear time invariant systems in the machine learning literature, that our estimation rates are explicitly \emph{tunable}. While the relationship between parameter identification and spectrum of input data is well known in the machine learning community \cite{Bishop2006, Vershynin2019}, it is worth stating that our framework provides a condition based on spectral lines by which a practitioner can achieve well-conditioned data in practice, subject to potential physical constraints, as outlined in Line 10 of Algorithm \ref{alg:example}. %In particular, the practitioner can achieve such well-conditioning by selecting an exogenous input with $n + m$ spectral lines, and our upper bound on estimation error can be optimized subject to any constraints as outlined in Line 10 of Algorithm \ref{alg:example}. 

For concreteness, we also provide two examples where the unmodeled dynamics are non-zero and estimation of the linearized dynamical system is still possible.
\begin{example}[High Pass Filter] \label{ex:high-pass}
Suppose the operator $f(\cdot)$ in \eqref{unmodeled-intro-w} is a high pass filter, such that
\[ \mathbf{w}(e^{jk/T}) = h(k/T) T \bar{\phi}(e^{jk/T}) \,, \]
where $h$ is an increasing function in frequency $k/T$. 

Suppose that for any $\epsilon > 0$, $h$ satisfies the property that there is a region of frequencies $[0, f_\epsilon]$ such that $h(f) \leq \epsilon$ for all $f \in [0, f_\epsilon]$. 
Then, for $T \geq \frac{n + m}{f_\ell}$, the practitioner may select $n+m$ frequencies such that $h(k/T) \leq \epsilon$.

Therefore, the estimation rates will take the form 
\[ 
    \max \left\lbrace ||\hat{A} - A_*||, || \hat{B} - B_*|| \right\rbrace \leq \tilde{O} \left( \sqrt{\frac{1}{T \norm{\bar{\Phi}^{-1}}^{-2}}} + \epsilon \right)\,.
\]
and since $\epsilon$ can be made arbitrarily small due to the assumption on the form of the transfer function of the unobserved dynamics, we see that consistent estimation rates can be achieved.
\end{example}
\begin{example}[Small Non-Linearities] \label{small-nonlinearity}
Suppose \eqref{unmodeled-intro-w} is such that

\begin{equation} \label{eq2}
\begin{aligned}
\overline w_{k} &= \sum_{i=1}^n a_i w_{k-i} + \sum_{i=1}^m b_i u_{k-i}\\
w_k&=f(\overline w_k) \,.
\end{aligned}
\end{equation}
If the function $f(\cdot)$ is such that the choice of $u_k$ will guarantee $\norm{w_k} = O(1/k^{1 + \epsilon})$ for some $\epsilon > 0$, then Parseval's theorem implies that 
\begin{align*}
\sum_{k = 0}^{T-1} \norm{\bar{\phi} (e^{jk/T}) \mathbf{w} (e^{jk/T})^\top } + \frac{1}{\sqrt{T}}  \norm{\mathbf{w} (e^{jk/T})} &\leq C \sum_{k = 0}^{T-1} \norm{\mathbf{w} (e^{jk/T})} = O(1) \,,
\end{align*}
as $\sum_{k = 1}^T \norm{w_k} = O(1)$. This results in consistent estimation. 
Such a condition can occur, for example, if $f(\cdot)$ corresponds to higher-order terms, and $u_k$ and initial conditions $w_0$ and $x_0$ are sufficiently small \cite{Khalil2002}.
\end{example}

\subsection{Lower Bound on Estimation with White Noise}
\label{s:lower-bound}
The previous theory indicates that the choice of an appropriate spectral lines-based input can result in consistent estimation. 
Here, we show that selecting the input as white noise can result in unwanted frequency response of the unmodeled dynamics.

To do so, we present the following general theorem for a lower bound, and then show an example which highlights the undesirable feature of white noise.

\begin{theorem} \label{lower-bound}
Consider the system \eqref{unmodeled-intro-x}-\eqref{unmodeled-intro-w} under Assumption \ref{wk-assumption}. 
Suppose control input $u_k$ is allowed to be stochastic and that the following condition holds for a fixed constant $\tau > 0$ and a probability $0 < \delta < 1$:
\begin{equation} \label{lower-bound-assumption}
    \P\left( \norm{\sum_{k = 0}^{T-1} \left(\frac{1}{T} \mathbf{u}(e^{jk/T}) \right) \left(\frac{1}{T} \mathbf{w}(e^{jk/T})^\top \right) }\geq \tau \right) \geq 1 - \delta/2.
\end{equation}

Then, for $(\hat{A}, \hat{B})$ as in \ref{least-squares}, with probability at least $1 - \delta$, for $T \gtrsim \tau^{-2}$,
\begin{align*}
     \max \left\lbrace \norm{\hat{A} - A_*}, \norm{ \hat{B} - B_*}\right\rbrace \geq c' \tau \,,
\end{align*}
where $c'$ is a constant with respect to $T$.
\end{theorem}

A proof of Theorem \ref{lower-bound} can be found in the Appendix.
The importance of Theorem \ref{lower-bound} is that as $T$ increases, the disparity in the estimation rates remains constant, such that consistent estimation is impossible under the condition \eqref{lower-bound-assumption}.
That is not to say that such a condition is impossible  for a spectral lines-based input.
Rather, for the deterministic input, it is the case that $\tau$ can be made arbitrarily small, as was the case in Example \ref{ex:high-pass}. 

The condition \eqref{lower-bound-assumption} encapsulates the idea that the exogenous input and the unmodeled dynamics are oscillating at the same frequencies with high probability, and is often satisfied when the exogenous input is selected to be white noise.
This occurs because white noise excites any frequency with high probability, and hence is likely to excite the frequencies present in the white noise.
In the following example, we show that when $u_k$ is selected as white noise, the magnitude of the parameter $\tau$ can not be controlled.
\begin{example}[High Pass Filter with White Noise Input] \label{ex:high-pass-wn}
Suppose, similar to Example \ref{ex:high-pass}, that the operator $f(\cdot)$ in \eqref{unmodeled-intro-w} is a high pass filter such that
\[ \mathbf{w}(e^{jk/T}) = h(k/T) \mathbf{u}(e^{jk/T}) \,, \]
where $h$ is an increasing function in frequency $k/T$. 

Suppose that for some $C > 0$, $h$ satisfies the property that there is a region of frequencies such that $h(f) \geq C$ for all $f \in [f_C, 1]$. 
Then, for any realization of the input $u_k$,
\begin{align*}
     \norm{\sum_{k = 0}^{T-1} \left(\frac{1}{T} \mathbf{u}(e^{jk/T}) \right) \left(\frac{1}{T} \mathbf{w}(e^{jk/T})^\top \right)} &= \frac{1}{T^2}\norm{\sum_{k = 0}^{T-1} h(k/T) \mathbf{u}(e^{jk/T}) \mathbf{u}(e^{jk/T})^\top} \\
     &\geq \frac{C}{T^2} \norm{\sum_{k = \lceil f_C T \rceil}^{T-1} \mathbf{u}(e^{jk/T}) \mathbf{u}(e^{jk/T})^\top} \\
     &\geq \frac{C}{n T^2} \textnormal{Tr} \left(\sum_{k = \lceil f_C T \rceil}^{T-1} \mathbf{u}(e^{jk/T}) \mathbf{u}(e^{jk/T})^\top \right) \\
     &= \frac{C}{n T^2} \sum_{k = \lceil f_C T \rceil}^{T-1} \norm{\mathbf{u}(e^{jk/T})}^2\,.
\end{align*}
If $u_k$ is taken to be white noise, then $\norm{\mathbf{u}(e^{jk/T})}^2$ will have an expectation which scales as $T$ \cite{Vershynin2019}, and the number of terms in the summation will also scale with $T$, so the expectation of the term $\frac{C}{T^2} \norm{\sum_{k = 0}^{T-1} h(k/T) \mathbf{u}(e^{jk/T}) \mathbf{u}(e^{jk/T})^\top}$ will exceed a constant.
Moreover, because $\norm{\mathbf{u}(e^{jk/T})}^2$ is the sum of a linear combination of squared Gaussian random variables, it will have the properties of a sub-Exponential distribution, and hence will concentrate around its expectation with high probability.
\end{example}

The primary difference between Examples \ref{ex:high-pass} and \ref{ex:high-pass-wn} lie in the ability to select the frequency content of the input.
In the deterministic setting, the frequency response of the unmodeled dynamics can be made arbitrarily small, oscillating at a constant number of frequencies with respect to $T$.
In contrast, for the stochastic input, the frequency response of the unmodeled dynamics can be large at a high number of frequencies which scales with $T$.
That is, because the white noise input does not allow for explicit frequency selection, this can result in a large magnitude of unobserved dynamics and hence inconsistent estimation.

We next turn to a key consequence of the spectral-lines based approach in terms of risk minimization for the LQR problem.

\section{Regret Bounds for Spectral Lines-Based Exploration}
\label{s:lqr-regret}
With the estimation results of Section \ref{s:unmodeled}, and the results of Section \ref{ss:regret_preliminaries} which are restated from Mania et al. \cite{Mania2019}, we are able to show that the spectral-lines based approach of Algorithm \ref{alg:example} results in $\Tilde{O}(\sqrt{T})$ regret for the LQR problem \eqref{eq:lqr}.
While we would wish to provide a guarantee on Regret for the general system \eqref{unmodeled-intro-x}-\eqref{unmodeled-intro-w}, this is not possible as the notion of an ``optimal'' controller in the general setting is ill-defined due to the nature of unmodeled dynamics.
That is, when the unmodeled dynamics $w_k$ in the system are non-zero, there is no general solution similar to the Ricatti equation which can be used for quadratic control.
Hence, we restrict our attention to the idealized case where dynamics are due to \eqref{lti-noisy-intro}, which is exactly the well-studied Linear Quadratic Regulator problem. 

Formally, our result is stated as follows:
\begin{theorem}\label{thm:regret}
Suppose Assumption \ref{assumption:strong-stability} holds for the system \ref{lti-noisy-intro}.
Then, with probability $1-\delta$, Algorithm \ref{alg:example} will achieve 
\begin{equation}
\textnormal{\textrm{Regret}}(T) \leq \tilde{O}(\sqrt{T}) \,.
\end{equation}
\end{theorem}
As in Theorem \ref{unmodeled-estimation}, the $\tilde{O}$ notation hides terms associated with dimensions $n$ and $m$, terms logarithmic in $T$ and $\frac{1}{\delta}$, and other parameters which are constant with respect to the underlying true parameters $(A_*, B_*, Q, R)$. 
Nevertheless, this result matches that of the machine learning literature which has shown a $\tilde{O}(\sqrt{T})$ regret for an algorithm similar to that of Algorithm \ref{alg:example} but with a Gaussian noise-based exploration \cite{Mania2019, Cohen2019}.
Here, we see that a deterministic spectral lines-based exploration meets the same regret bound.
A proof of Theorem \ref{thm:regret} can be found in the Appendix.

It is worth noting that this result, which is meaningful for sufficiently large $T$, still requires a stabilizing controller for the first epoch, and that the estimates $(\hat{A}, \hat{B})$ from the \emph{first} estimation procedure must satisfy $\max\{\norm{\hat{A} - A_*}, \norm{\hat{B} - B_*}\} \leq \epsilon (A_*, B_*, Q, R)$ where $\epsilon(A_*, B_*, Q, R)$ is a sufficiently small value determined by the underlying true system \cite{Mania2019}.
Such accurate estimates can be guaranteed by ensuring that the length of the first epoch, given by the constant $C(A_*, B_*, Q, R, \delta)$ in Line 3 of Algorithm \ref{alg:example} is sufficiently large, as Theorem \ref{unmodeled-estimation} will then ensure accurate estimates with high probability.
That is, because the result of the previous literature is local in nature, the constants associated with the theorem statement here may be quite large in practice.

However, the spectral-lines based method still provides reasonable risk minimization in simulation.
In the following section, we provide experiments which indicate the spectral-lines based approach minimizes risk even for small $T$.
Moreover, our experimental results suggest that the approach of Algorithm \ref{alg:example} is general in the sense that it can be applied to the setting of \eqref{unmodeled-intro-x}-\eqref{unmodeled-intro-w} with better performance than that of the white-noise based approach considered in previous literature.

\section{Experimental Evaluation}
\label{s:experiments}
Reliable autonomous systems must account for the uncertainty inherent to the tasks which they aim to solve.
Our theoretical results above indicate that, by using a spectral lines-based approach, one can reliably estimate the parameters of a dynamical system even in the case where deterministic unmodeled dynamics can be present, and have the potential to outperform a Gaussian-noise based input in this task.
The results above hold for large $T$, which may not always be feasible for the data collection of a risk-sensitive autonomous system.
In practice, we still expect the spectral-lines based approach to outperform the approach based on noise, even in the regime with small data.

To this end, we provide experimental evidence which illustrates the efficacy of a spectral lines-based approach in estimating parameters and minimizing risk for a dynamical system.
We show these results in two parts.
First, we highlight the performance of the spectral lines-based approach in estimation of a system of the form \eqref{unmodeled-intro-x}-\eqref{unmodeled-intro-w}, and show that the proper choice of frequencies for spectral lines can lead to significant improvements in parameter estimation.
Next, we show how the use of spectral lines-based exploration in the setting of the Linear Quadratic Regulator can be used achieve low regret in practice.
Our results indicate that while the spectral lines-based approach is comparable to a Gaussian noise-based approach in the system \eqref{lti-noisy-intro}, it outperforms the noise-based approach for the system \eqref{unmodeled-intro-x}-\eqref{unmodeled-intro-w} where the noise in the input can disrupt the system dynamics in unexpected ways.

For both experiments, for simplicity we will use the following structure on the dynamical system.
In both models, we assume the following parameterization of $A_*$ and $B_*$:
\begin{align}
    A_* = \begin{bmatrix}
    0 & 1 & 0 \\
    0 & 0 & 1 \\
    a_1 & a_2 & a_3
    \end{bmatrix}, \qquad B_* = \begin{bmatrix}
    0 \\ 0 \\ 1
    \end{bmatrix} \label{eq:matrices}
\end{align}
As such, these linear components are in control canonical form \cite{Goodwin_1984, Ljung_1987}.
This parameterization of the linear dynamics can be done without loss of generality, in the sense that invertible transformations of the observations can always be made such that the system has the form \eqref{eq:matrices}.
Moreover, this parameterization allows for the spectrum of $A_*$ to be selected in a simple way for each problem setting, as the spectrum will only depend on the three parameters $a_1$, $a_2$, and $a_3$.
The unmodeled dynamics are represented by the following non-linear dynamics:
\begin{align}
    w_{k} &= \begin{bmatrix} 1 & 1 & 1
    \end{bmatrix}^\top \times c \times \overline{w}_k^2, \label{eq:unmodeled-nonlin} \\ \overline{w}_k &= \alpha \overline{w}_{k-1} + \alpha(u_k - \beta u_{k-1}) \label{eq:unmodeled-highpass}
\end{align}
which represent a non-linear transformation in the form of a high pass filter on the control input $u_k$. 
For the simulation results, we use the system defined in \eqref{eq:matrices}-\eqref{eq:unmodeled-highpass} to represent the system \eqref{unmodeled-intro-x}-\eqref{unmodeled-intro-w}.
This choice of parameterization is meant to reflect realistic scenarios present in physical systems, wherein local dynamics are driven mostly by a linear system, but a high frequencies can disrupt the system in unexpected ways.

\subsection{Parameter Estimation with Unmodeled Dynamics}

To approach the task of parameter estimation, we use two types of control inputs in order to estimate the system. 
The first is a white noise-based input, as in Sarkar and Rakhlin \cite{Sarkar_2019}, and is selected as
\begin{equation}
    \label{eq:noise_input}
    u_k = s_k \qquad s_k \sim \mathcal{N}(0, E_0^2) \,,
\end{equation}
The second, a spectral lines-based input, is selected as
\begin{equation}
    \label{eq:pe_input}
    u_k = u_{spec, k} = E_0' \sum_{i = 1}^2 \cos (2 \pi f_i t) \,,
\end{equation}
where $E_0'$ is such that the selection of control inputs ensures $\sum_{k = 1}^T \E\left[ s_k^2 \right] = \sum_{k = 1}^T u_{spec,k}^2 = T E_0$. 
This choice is made to provide a comparison of signals with equivalent energy. 
We first compare the effect of our proposed approach in \eqref{eq:pe_input} with that in \eqref{eq:noise_input} based on parameter estimation, and discuss results based on Regret in the following subsection. 
For each setting, the frequencies $f_i$ are selected based on the problem setting, emulating how practitioners may select frequency content based on the application. 
Details and parameter selections are discussed in Appendix H.

\begin{figure}[!b]
    \centering
    \includegraphics[width = 0.7\textwidth]{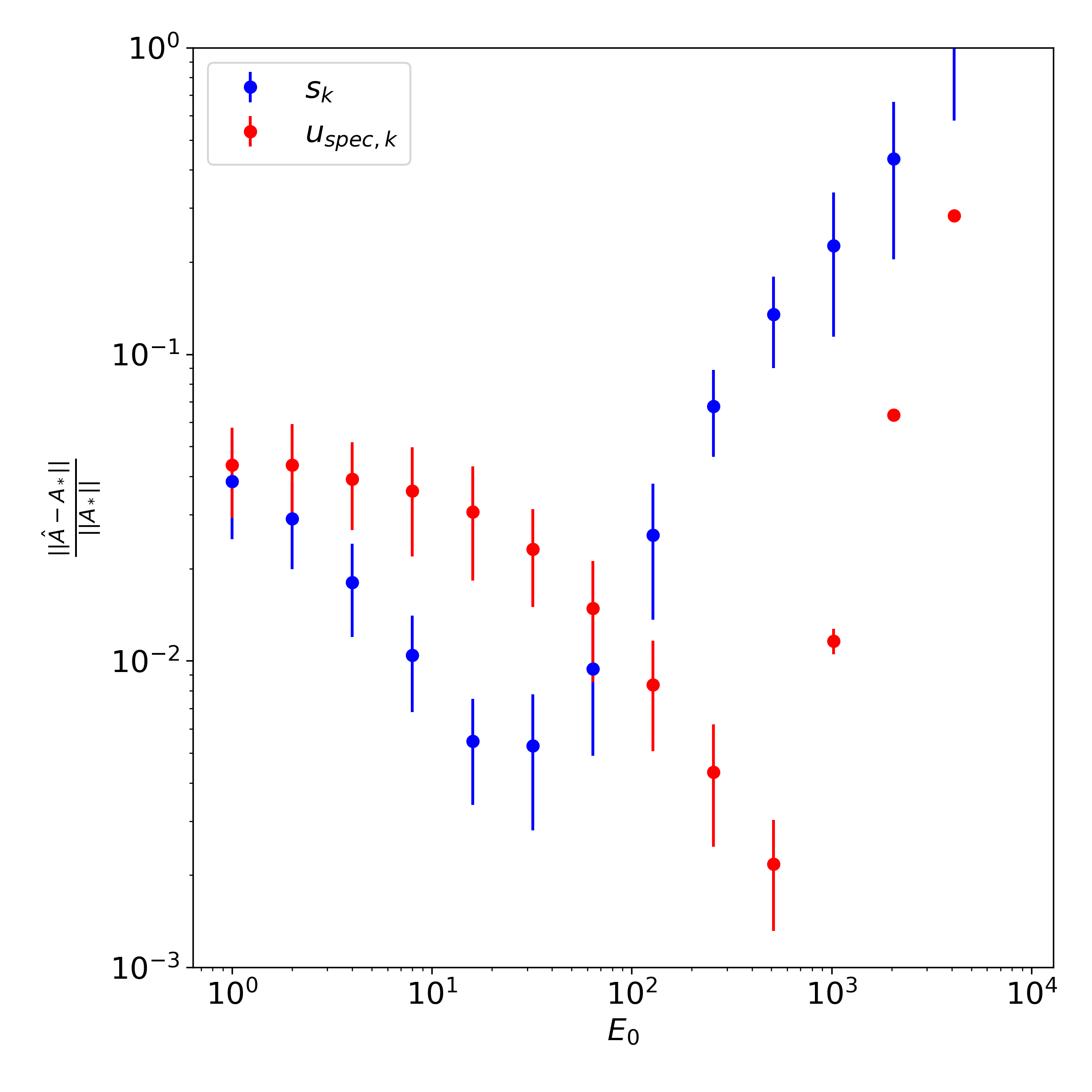}
    \caption{Estimation errors for the system \eqref{eq:matrices}-\eqref{eq:unmodeled-highpass}, for various levels $E_0$ of energy of input signals. 
    Frequencies for the spectral-lines based input are $f_1 = 0.01$ and $f_2 = 0.05$, which are selected to be low values due to the knowledge that the unmodeled dynamics reflect a high pass filter. 
    Experiments are run $100$ times, for $T = 500$, and error bars correspond to one standard deviation.
    We see that across all considered choices of the exogeneous input, the lowest estimation error is achieved by the spectral lines-based approach (when $E_0 = 500$), and that at high energy levels the spectral lines-based method consistently outperforms the white noise-based method.}
    \label{fig:exp-results}
\end{figure}

Figure \ref{fig:exp-results} illustrates the estimation errors obtained using \eqref{eq:pe_input} (red) and \eqref{eq:noise_input} (blue) as a function of the input energy $E_0$.
Notably, at higher energies, the spectral-lines based input has both lower estimation error and less variability in the parameter estimation error, as the spectral-lines based input does not excite the unmodeled dynamics.
The experiments also reflect a non-trivial relationship between the energy of the input and the estimation error, as higher energy does not always lead to better estimation when unmodeled dynamics are present.
Rather, in contrast to previous literature which suggests higher levels of exploration noise leads to tighter estimation rates, here we instead see that there is a point at which increased exploration magnitude results in poor estimation due to the exacerbation of the unmodeled dynamics.
Such a point happens for lower levels of energy for the noise-based input, and hence more accurate estimation can be achieved with the spectral lines-based input.

\subsection{Regret of the Linear Quadratic Regulator}

\begin{figure}[!t]
    \centering
    \includegraphics[trim=0 0 0 10, width = 0.7\textwidth]{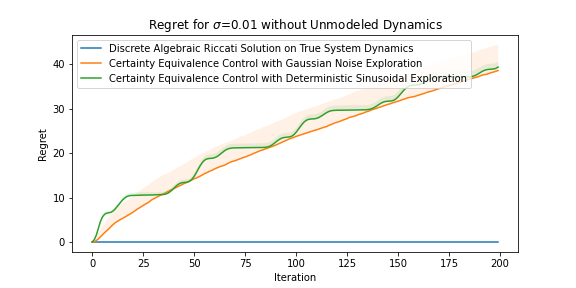}
    \includegraphics[trim=0 10 0 0, width = 0.7\textwidth]{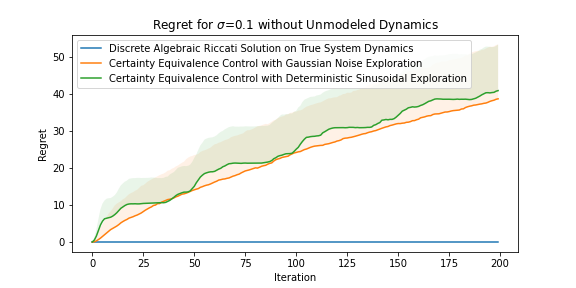}
    \caption{Median and 90th percentile of regret over time, at two levels of external noise, with no unmodeled dynamics. In both cases, the spectral lines-based exploration (with $f_1 = 0.03$ and $f_2 = 0.05$) has slightly higher median regret than the Gaussian noise-based exploration, but the regrets for both grow at the same rate.}
    \label{fig:regret-results}
\end{figure}

To demonstrate the efficacy of spectral lines-based input with respect to risk minimization of an autonomous system, we apply Algorithm \ref{alg:example} in simulation. 
We first simulate the idealized system in \eqref{lti-noisy-intro}, implemented as in \eqref{eq:matrices} with eigenvalues at 0.9959, 1.01, and 1.0241, which has been considered in recent literature and for which we are able to provide theoretical guarantees on regret.
Figure \ref{fig:regret-results} shows the regret from LQR control of Algorithm \ref{alg:example} as compared to the algorithm for the LQR used in Mania et al. \cite{Mania2019}, which parallels Algorithm \ref{alg:example} except in that the exploration $u_{spec, k}$ is replaced by independent realizations of Gaussian noise.
The results indicate that for the ideal system in \eqref{lti-noisy-intro}, regret is comparable for both approaches.

\begin{figure}[!t]
    \centering
    \includegraphics[trim=0 0 0 10, width = 0.7\textwidth]{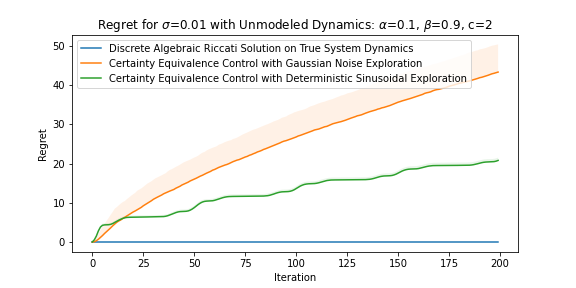}
    \includegraphics[trim=0 10 0 0, width = 0.7\textwidth]{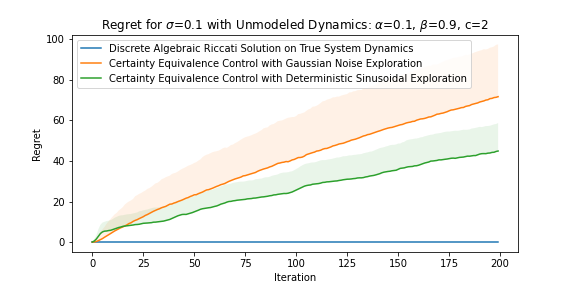}
    
    \caption{Median and 90th percentile of Regret over time, at two levels of external noise, with unmodeled dynamics present. In both cases, the spectral lines-based exploration (with $f_1 = 0.03$ and $f_2 = 0.05$) outperforms the Gaussian noise-based exploration.}
    \label{fig:regret-results-unmodeled}
\end{figure}

However, when unmodeled deterministic dynamics are present in the system, as they often are in physical systems, we see that the spectral-lines based exploration outperforms Gaussian noise-based exploration.
Figure \ref{fig:regret-results-unmodeled} highlights this difference, as it shows the regret from LQR control of a more realistic system defined in \eqref{eq:matrices}-\eqref{eq:unmodeled-highpass}. 
As in the previous example, performance is compared between Algorithm \ref{alg:example} as presented, and Algorithm \ref{alg:example} with the modification that $u_{spec, k}$ in line 6 of Algorithm 1 is replaced with $s_k$ from \eqref{eq:noise_input}.
In both experiments, the baseline for calculating regret is an ``optimal'' controller for which $A_*$ and $B_*$ are known and lines 6 and 7 in Algorithm 1 are replaced by $u_k = K_*x_k$, where $K_*$ is the solution to the discrete algebraic Riccati equation. The 

In both simulations, unobserved stochastic noise is present, i.e. $\eta_t$ in \eqref{lti-noisy-intro} and \eqref{unmodeled-intro-x}-\eqref{unmodeled-intro-w} has strictly positive variance.
Each figure contains the results for two noise levels: the top plots use a noise-to-signal ratio of $\sigma = 0.01$, typical in practical applications \cite[Chapter 6.2.1]{Lavretsky2013}, and the bottom plots use a noise-to-signal ratio of $\sigma = 0.1$, to compare to a situation in which noise is larger. 
Additionally, in Figure \ref{fig:regret-results-unmodeled}, the parameters in \eqref{eq:unmodeled-nonlin}-\eqref{eq:unmodeled-highpass} are chosen so that the magnitude of unmodeled dynamics is approximately an order of magnitude smaller than the modeled part at low frequencies.
Figure \ref{fig:regret-results} shows that without unmodeled dynamics, our approach performs comparably to noise-based exploration.
This is unsurprising as there are no high-frequency dynamics for the Gaussian noise to excite. 
However, as discussed in Section 1, unmodeled dynamics are always present for physical systems. 
Figure \ref{fig:regret-results-unmodeled}, then, shows that in a more realistic setting, our approach improves upon the performance of noise-based exploration. 
Additionally, it is worth noting that in our approach, the only noise into the system is the unobserved noise $\eta_k$. Therefore, it is visible in both figures that when the noise-to-signal ratio is small (i.e. in the $\sigma = 0.01$ case), the spectral lines-based exploration has significantly decreased variability in its regret.

\subsection{Regret of LQR on Larger Systems}

A larger LTI system ($n = 5$) was also simulated to show the efficacy of our approach over Gaussian noise-based exploration on other systems. The new system took the form of \eqref{eq:matrices-larger}-\eqref{eq:unmodeled-highpass-larger}. For comparability, the eigenvalues were placed at 0.9959, 1.003, 1.01, 1.0171, and 1.0241, at or near the eigenvalues of the system simulated in Section 7.2.
\begin{align}
    A_* &= \begin{bmatrix}
    0 & 1 & 0 & 0 & 0 \\
    0 & 0 & 1 & 0 & 0 \\
    0 & 0 & 0 & 1 & 0 \\
    0 & 0 & 0 & 0 & 1 \\
    a_1 & a_2 & a_3 & a_4 & a_5
    \end{bmatrix}, \qquad B_* = \begin{bmatrix}
    0 \\ 0 \\ 0 \\ 0 \\ 1
    \end{bmatrix} \label{eq:matrices-larger}
\end{align}
\begin{align}
    w_{k} &= \begin{bmatrix} 1 & 1 & 1 & 1 & 1
    \end{bmatrix}^\top \times c \times \overline{w}_k^2, \label{eq:unmodeled-nonlin-larger} \\ \overline{w}_k &= \alpha \overline{w}_{k-1} + \alpha(u_k - \beta u_{k-1}) \label{eq:unmodeled-highpass-larger}
\end{align}
\begin{figure}[!ht]
    \centering
    \includegraphics[trim=0 0 0 10, width = 0.7\textwidth]{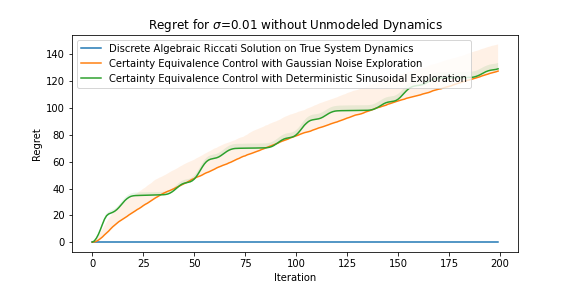}
    \includegraphics[trim=0 10 0 0, width = 0.7\textwidth]{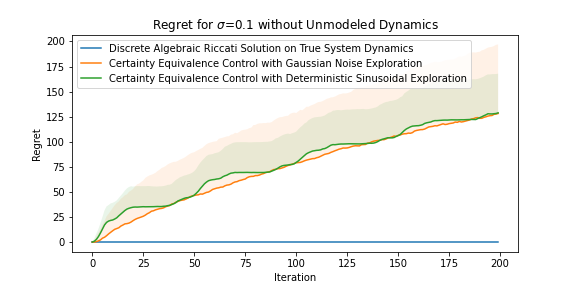}
    
    \caption{Simulations of a larger LTI system ($n = 5$). Median and 90th percentile of Regret over time, at two levels of external noise, with no unmodeled dynamics. In both cases, the spectral lines-based exploration with ($f_1 = 0.03$, $f_2 = 0.05$, and $f_3 = 0.07$) has slightly higher median regret than the Gaussian noise-based exploration, but the regrets for both grow at the same rate.}
    \label{fig:regret-results-larger}
\end{figure}

\begin{figure}[!t]
    \centering
    \includegraphics[trim=0 0 0 10, width = 0.7\textwidth]{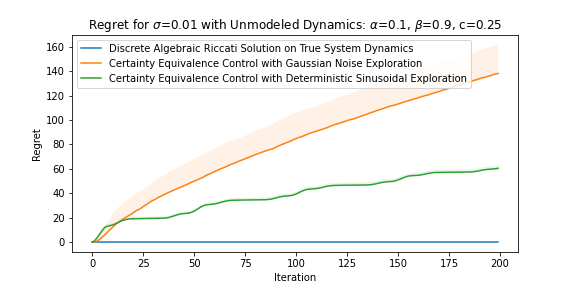}
    \includegraphics[trim=0 10 0 0, width = 0.7\textwidth]{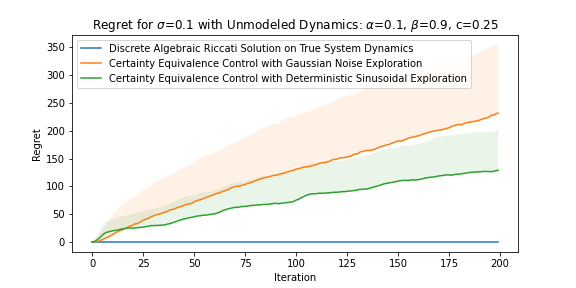}
    
    \caption{Simulations of a larger LTI system ($n = 5$). Median and 90th percentile of Regret over time, at two levels of external noise, with unmodeled dynamics present. In both cases, the spectral lines-based exploration with ($f_1 = 0.03$, $f_2 = 0.05$, and $f_3 = 0.07$) outperforms the Gaussian noise-based exploration.}
    \label{fig:regret-results-unmodeled-larger}
\end{figure}

Figure \ref{fig:regret-results-larger} shows the simulation results for the ideal case with no unmodeled dynamics present, i.e. \eqref{lti-noisy-intro} implemented as \eqref{eq:matrices-larger}, and Figure \ref{fig:regret-results-unmodeled-larger} shows the simulation results for the more realistic case with unmodeled dynamics given in \eqref{eq:matrices-larger}-\eqref{eq:unmodeled-highpass-larger}. As in Section 7.2, the top plot of each figure uses a noise-to-signal ratio of 0.01 and the bottom plot uses a ratio of 0.1. The results in Figure \ref{fig:regret-results-larger} show that as in the $n = 3$ case, a deterministic persistently exciting input is comparable to a Gaussian noise-based input without unmodeled dynamics present. Finally, Figure \ref{fig:regret-results-unmodeled-larger} demonstrates that, just as in the $n = 3$ case, our approach outperforms a Gaussian noise-based input when unmodeled dynamics are present.

\section{Conclusions}
\label{s:conclusions}
In this paper, we propose the use of a spectral lines-based deterministic exogenous signal to carry out parameter estimation in physical dynamic models where deterministic unmodeled dynamics are
almost always present. Our theoretical analysis
consists of a new toolkit which employs the theory
of spectral lines, retains the stochastic setting,
and helps derive non-asymptotic bounds on the parameter estimation error. The results are shown to lead to a tunable parameter identification error. In particular, it is shown that the identification error can be minimized through an optimal choice of the spectrum of the exogenous signal. 

Experiments are provided which show the efficacy of a spectral lines-based approach in comparison to the noise-based approach both in terms of parameter estimation error as well as Regret obtained using the parameter estimates in a closed-loop control
design using the Linear Quadratic Regulator
method. In particular, these experiments show that for the same level of input energy, our method can lead to over a 59\% decrease in the parameter estimation error compared to the noise-based method. Similarly, for a given set of unmodeled dynamics, the Regret achieved by the Linear Quadratic Regulator is halved after 200 iterations when using the spectral lines-based method over the white noise based method.

%\newpage
% \section*{Broader Impact}
% \input{body/broad-impact.tex}

\section*{Acknowledgements and Disclosure of Funding}
This work was funded in part by the Boeing Strategic University Initiative. The first author gratefully acknowledges a Graduate Fellowship through the MIT Institute for Data, Systems, and Society.

\bibliography{main}
\newpage
{\Large \bf Appendix}
\appendix
\paragraph{Organization of the appendix.}

We discuss the continuous time analogue to Section \ref{ss:adaptive_preliminaries} in \ref{a:continuous-background}, to help inform the reader of well-studied contexts for parameter convergence discussed in the adaptive control literature. Probabilistic inequalities are provided in \ref{a:prob-inequalities}, which are used in Appendices \ref{a:transfer-proof}, \ref{a:s-to-pe},\ref{a:unmodeled-proof-upper}, and \ref{a:unmodeled-proof-lower} to provide detailed proofs of claims. \ref{a:experiment-details} discusses specific details and parameter selections with respect to the experiments shown in the paper.

\section{Time-Varying Regression in Continuous Time}
\label{a:continuous-background}
\subsection{Parameter identification in a regression model}

In this section, we provide a brief introduction to regression with time-varying regressors in continuous time. Consider a regression system of the form
\begin{equation}
\label{aa0}
y(t)=\theta^{*\top}\phi(t),
\end{equation}
where $\theta^*\in\mathbb{R}^N$ represents an unknown constant parameter and $\phi:\mathbb{R}^+\rightarrow\mathbb{R}^N$ represents a known time-varying regressor. The variable $y:\mathbb{R}^+\rightarrow\mathbb{R}^N$ represents a known time-varying output. Given that $\theta^*$ is unknown, we formulate an estimator $\hat{y}(t)=\theta^\top(t)\phi(t)$, where $\hat{y}:\mathbb{R}^+\rightarrow\mathbb{R}$ is the predicted output and the unknown parameter is estimated as $\theta$, where $\theta:\mathbb{R}^+\rightarrow\mathbb{R}^N$. This leads to an error model of the form
\begin{equation}
\label{e:error1}
e_y(t)=\tilde{\theta}^\top(t)\phi(t),
\end{equation}
where $e_y=\hat{y}-y$ is an output error, and $\tilde{\theta}=\theta-\theta^*$ is the parameter estimation error and the two errors are related through (\ref{e:error1}). The primary goal is to design a rule to adjust the parameter estimate $\theta$ in a continuous manner using knowledge of $\phi$ and $e_y$ such that the unknown parameter is estimated, i.e., $\tilde\theta(t)\rightarrow 0$ as $t\rightarrow \infty$. The secondary goal is to at least ensure that the adjustment rule provides for the convergence of the output error $e_y(t)$ towards zero. A gradient-flow based algorithm is often suggested for this purpose \cite{Ioannou1996}. A squared output error loss function $L=(1/2)e_y^2$ is often used to lead to an update rule
\begin{equation}
\label{e:update_GF}
\dot{\theta}(t)=-\gamma \phi(t)e_y(t),
\end{equation}
where $\gamma>0$ is a user defined gain.

A careful application of stability theory easily guarantees that $\theta(t)$, the solutions of (\ref{e:update_GF}), are bounded for any initial condition $\theta(t_0)$. For any bounded regressor $\phi(t)$ with a bounded time derivative, it can also be shown that the output error converges to zero, i.e., that the secondary goal is achieved. This raises the question as to when the primary goal of accurate parameter estimation is achieved. This problem is equivalent to determining the asymptotic stability of the equilibrium point $\tilde{\theta}(t)=0$ of the time-varying differential equation
\begin{equation}
\label{aa1}
\dot{\tilde{\theta}}(t)=-\gamma\phi(t)\phi^\top(t)\tilde{\theta}(t).
\end{equation}
For this, we need the regressor to satisfy an additional condition known as persistent excitation (PE). This condition is expanded upon in the following section.

\subsection{Persistent Excitation in Continuous Time}
\begin{definition}\label{d:PE}
	\cite{Morgan_1977,Morgan_1977a} A function $\phi:\R^+\rightarrow \R^N$ is persistently exciting (PE) if there exists $T>0$ and $\alpha>0$, $\beta>0$ such that
	\begin{equation}\label{PE}
	\beta I\geq\int_t^{t+T}\phi(\tau)\phi^\top(\tau)d\tau\geq\alpha I,\quad t\geq t_0.
	\end{equation}
\end{definition}
\begin{theorem}\label{th:PE}
\cite{Morgan_1977,Morgan_1977a} A persistently exciting regressor $\phi$ is necessary and sufficient for the uniform asymptotic stability of $\tilde\theta=0$ in (\ref{aa1}).
\end{theorem}
 
We note that the preceding discussion revolves around deterministic signals. This together with the fact that the stability property in Theorem \ref{th:PE} is uniform and asymptotic imply that given an $\epsilon>0$, there exists a finite $T>0$ such that 
\begin{equation}\label{conv}
\norm{\tilde\theta(t)}\leq \epsilon, \qquad \forall t\geq t_0+T.
\end{equation}
We refer the reader to the original sources \cite{Morgan_1977,Morgan_1977a} for the details of the proof of Theorem \ref{th:PE}.

A spectral lines-based necessary and sufficient condition for a regressor to satisfy the persistent excitation condition in Definition \ref{d:PE} may alternatively be provided, as discussed in \cite{Boyd_1986, Boyd_1983}. Before proceeding to this theorem, we recall the definition of a spectral line.

\begin{definition}\cite{Boyd_1983}
A function $u:\R^+\rightarrow\R^n$ is said to have a spectral line at a frequency $\omega_0$ of amplitude $\bar{u}(\omega_0)\in \C^n$ iff
\begin{equation}
\frac{1}{T}\int_t^{t+T} u(\tau)e^{-j\omega_0\tau}d\tau
\label{spectral}
\end{equation}
converges to $\bar{u}(\omega_0)$ as $T\rightarrow \infty$ uniformly in $t$.
\end{definition}

The following theorem uses the definition of a spectral line to relate the persistent excitation condition in Definition \ref{d:PE} to spectral lines of an input of a dynamical system.

\begin{theorem}\cite{Boyd_1986}
Suppose that $\phi$ is given by the state of a linear dynamic system,
\[ \dot{\phi} = A \phi + b r(t) \,, \]
where $(A,b)$ is controllable, $A \in \R^{N \times N}$ and $b \in \R^N$. Then, $\phi$ satisfies the PE condition unless the spectrum of $r(t)$ is concentrated on $k<N$ lines.
\end{theorem}

\section{Probabilistic Inequalities}
\label{a:prob-inequalities}

\begin{proposition}[Sum of Dependent sub-Gaussian Random Variables] \label{sum-dependent}
Let $X \sim \subG(\sigma^2)$ and $Y\sim\subG(\tau^2)$ be two arbitrarily dependent sub-Gaussian random variables. Then,
\[ X + Y \sim \subG((\sigma + \tau)^2) \,. \]
\end{proposition}
\begin{proof}
By H\"older's inequality, 
\begin{align*}
   \E[e^{\lambda(X + Y)}] &\leq \left(\E[e^{p\lambda X}] \right)^{1/p} \left(\E[e^{q\lambda Y}] \right)^{1/q} &&\left(\frac{1}{q} + \frac{1}{p} = 1\right) \\
   &\leq e^{\frac{\lambda^2}{2} \left( \frac{q \sigma^2}{q-1} + q \tau^2 \right)} \,.
\end{align*}
The claim then follows by setting $q = \frac{\sigma}{\tau} + 1$.
\end{proof}

In particular, by induction, the above proposition states that for random variables $X_1, \dots, X_n$, where $X_i \sim \subG(\sigma_i^2)$, $X_1 + \dots + X_n \sim \subG\left( \left(\sum_{i=1}^n \sigma_i \right)^2 \right)$.

\begin{proposition}[Operator Norm with Dependent sub-Gaussian Entries] \label{op-norm-dependent}
Let $M \in \R^{n \times n}$ be a random matrix with dependent sub-Gaussian entries with variance proxy $R^2$. Then,
\[ \P(\norm{M} > t) \leq 9^{2n} e^{-\frac{c t^2}{n R^2}} \,,\]
for a universal constant $c$.
\end{proposition}
\begin{proof}
First, we note
\[ \norm{M} = \sup_{x \in \cS^{n-1}, y \in \cS^{n-1}} x^\top M y \,.\]
Let $\mathcal{E}^n$ be a $1/4$-net of $\cS^{n-1}$, which has size at most $9^n$ \cite{Vershynin2019}. Then, we see:
\[ \norm{M} \leq 2 \sup_{w \in \mathcal{E}^n, z \in \mathcal{E}^n } w^\top M z \]
Hence, for arbitrary $w, z \in \mathcal{E}^n$, we see
\begin{align*}
    \P(\norm{M} > t) \leq 9^{2n} \P\left( w^\top W z > \frac{t}{2} \right) \,.
\end{align*}
Finally, we may note that $w^\top M z \sim \subG(n R^2)$. This follows from Proposition \ref{sum-dependent} and the fact that $\sum_{i, j} |w_i| |z_j| \leq n$ when $w, z \in \cS^{n-1}$. Therefore, for an absolute constant $c$,
\[ \P(\norm{M} > t) \leq 9^{2n} e^{-\frac{c t^2 }{n R^2}} \,, \]
as desired.
\end{proof}

\begin{proposition}[\cite{Vershynin2019}] \label{random-sup-norm}
Let $M \in \R^{n \times d}$ be a random matrix. Then, for any $\epsilon < 1$ and any $w \in \cS^{d-1}$, 
\[ \P(\norm{M} > z) \leq \left(1 + \frac{2}{\epsilon} \right)^d \P(\norm{Mw} > (1 - \epsilon) z)\]
\end{proposition}
A proof of the above claim may be found in Vershynin \cite{Vershynin2019}. 

Similar to Sarkar and Rakhlin \cite{Sarkar_2019}, we will be able to use the above claim to help us analyze self-normalized martingale terms, as initially discussed in Abbasi-Yadkori et al. \cite{Abbasi2011} with the following theorem. We will also be able to apply similar arguments to the above proposition to the setting of random matrices with arbitrarily dependent sub-Gaussian entries.

\begin{theorem}[Theorem 1 in Abbasi-Yadkori et al. \cite{Abbasi2011}] \label{th:MartingaleTheorem}
Let $\{\cF_k \}_{k=1}^\infty$ be a filtration. Let $\{\eta_k\}_{k=1}^\infty$ be a sequence of real-valued random variables such that $\eta_k$ is $\cF_{k+1}$-measurable and conditionally $R$ sub-Gaussian given $\cF_{k}$, i.e.
\[ \E[e^{\lambda\eta_k} | \cF_{k}] \leq e^{\frac{\lambda^2 R^2}{2}} \qquad \forall~\lambda > 0 \,. \]
Let $\{ x_k \}_{k=1}^\infty$ be a sequence of random vectors such that $x_k \in \R^d$ is $\cF_k$ measurable, and let $V \in \R^{d \times d}$ be an arbitrary positive definite deterministic matrix, and define
\[ \bar{V}_T = V + \sum_{k = 1}^T x_k x_k^\top, \qquad S_t = \sum_{k = 1}^T x_k \eta_k \,. \]
Then, for any $\delta > 0$, with probability at least $1 - \delta$, for all $T > 0$, 
\[ \norm{S_T}_{\bar{V}_T}^2 \leq 2R^2 \log \left( \frac{\det(\bar{V}_T)^{1/2} \det(V)^{-1/2}}{\delta} \right)\]
\end{theorem}

Using the above claim, we now restate Proposition \ref{self-norm-body} in the following proposition, complete with proof.

\begin{proposition}[Self-Normalized Martingale Bound]
Consider the system \eqref{unmodeled-intro-x}-\eqref{unmodeled-intro-w} where $w_k, \eta_k, x_k \in \R^n$, $\eta_k \sim \subG(\sigma^2)$, and $u_k \in \R^m$ are deterministic. For an arbitrary deterministic matrix $V \succ 0$, define
\[ \phi_k = \begin{bmatrix}
x_k \\ u_k
\end{bmatrix}, \qquad \bar{Y}_T = V + \sum_{k=1}^T \phi_k \phi_k^\top, \qquad S_T = \sum_{k = 1}^T \phi_k \eta_k^\top. \]
Then, for any $0< \delta < 1$, with probability at least $1 - \delta$,
\[ \norm{(\bar{Y}_T)^{-1/2} S_T} \leq \sigma \sqrt{8(n+m) \log \left( \frac{5 \det(\bar{Y}_T)^{1/(2(n+m))} \det(V)^{-1/(2(n+m))}}{\delta^{1/(n+m)}} \right)} \,. \]
\end{proposition}
\begin{proof}
From Proposition \ref{random-sup-norm}, setting $\epsilon = 1/2$, we note that for any $y$,
\begin{align*}
    \P \left(\norm{(\bar{Y}_T)^{-1/2} S_T} > y \right) &\leq 5^{n+m} \P \left(\norm{(\bar{Y}_T)^{-1/2} S_T w} > \frac{y}{2} \right) \\
    &= 5^{n+m} \P \left(\norm{(\bar{Y}_T)^{-1/2} S_T w}^2 > \frac{y^2}{4} \right) \,.
\end{align*}
Note that $S_T w = \sum_{k = 1}^T \phi_k \eta_k^\top w$, and that $\eta_k^\top w \sim \subG(R^2)$ by the definition of a sub-Gaussian random vector. Letting $\cF_k$ be the $\sigma$-algebra generated by  $(x_k, \dots, x_0, \eta_{k-1}, \dots, \eta_{0})$, we are then able to apply Theorem \ref{self-norm-body}, where $\eta_k^\top w$ now corresponds to $\eta_k$, and the claim follows by setting
\[ y^2 = 8\sigma^2 \log \left( \frac{ \det(\bar{Y}_T)^{1/2} \det(V)^{-1/2}}{5^{-(n+m)}\delta} \right) \,, \]
as this choice will ensure $\P \left(\norm{(\bar{Y}_T)^{-1/2} S_T w}^2 > \frac{y^2}{4} \right)\leq 5^{-(n+m)} \delta$.
\end{proof}

\begin{remark}
    The above proposition may then be applied to an arbitrary system, assuming there is a deterministic upper bound on $\bar{Y}_T$ with high probability. We will show this indeed the case for both \eqref{lti-noisy-intro} and \eqref{unmodeled-intro-x}-\eqref{unmodeled-intro-w} with the following results.
\end{remark}

\begin{proposition}[Markov's Inequality \cite{Vershynin2019}] \label{markov-inequality}
Consider an integrable random variable $X$ defined on a probability space $(\Omega, \cF, \P)$ such that for all $\omega \in \Omega$, $X(\omega) \geq 0$. Then, for any $\delta > 0$,
\[ \P\left(X \geq \frac{\E[X]}{\delta} \right) \leq \delta \,.\]
\end{proposition}
\begin{proof}
For all $\omega \in \Omega$, and any $a$, note $X(\omega) \geq a \1_{X(\omega) \geq a}$. Thus,
\[ \E[X] \geq \E[a \1_{X(\omega) \geq a}] = a \P(X \geq a) \,.\]
The claim follows by rearranging the inequality above and setting $a = \E[X]/\delta$.
\end{proof}

In order to show the upper bounds on $\bar{Y}_T$, we must also define the following two system-dependent matrices.

\begin{definition}[Gramian Matrix]\label{gramian}
For a matrix $A \in \R^{n \times n}$, the Gramian of $A$ is defined
\[ \Gamma_k (A) = \sum_{i = 0}^k A^i (A^i)^\top \,. \]
\end{definition}

\begin{definition}[Controllability Gramian]\label{controllability-gramian}
For matrices $A \in \R^{n \times n}$ and $B \in \R^{n \times m}$, the Controllability Gramian of the pair $(A,B)$ is defined
\[ \Gamma_k (A, B) = \sum_{i = 0}^k A^i B B^\top (A^i)^\top \,. \]
\end{definition}

\begin{proposition}[Deterministic Upper Bound for LTI Systems] \label{vup}
Consider the system
\[ x_{k+1} = A_* x_k + B_* u_k + \eta_k \,, \]
where $\eta_k, x_k \in \R^n$, $\eta_k \sim \subG(\sigma^2)$, and $u_k \in \R^m$ are deterministic for each $k$. Define
\[ \phi_k = \begin{bmatrix}
x_k \\ u_k
\end{bmatrix}, \qquad Y_T = \sum_{k=0}^{T-1} \phi_k \phi_k^\top \,. \]
Then, for any $\delta > 0$, with probability at least $1 - \delta$,
\[  Y_T \preceq \left( \frac{\sigma^2 T \tr(\Gamma_{T-1}(A_*)) + T u_M^2 \tr(\Gamma_{T-1} (A_*,B_*)) + Tu_M^2}{\delta} \right) I \,. \]
\end{proposition}
\begin{proof}
We first define the quantities
\[ \tilde{A} = \begin{bmatrix}
I & 0 & \dots & 0 \\
A_* & I & \dots & 0 \\
\vdots & \vdots & \ddots & \vdots \\
A_*^{T-1} & A_*^{T-2} & \dots & I 
\end{bmatrix}, \qquad \tilde{B} = \begin{bmatrix}
B_* & 0 & \dots & 0 \\
A_*B_* & B_* & \dots & 0 \\
\vdots & \vdots & \ddots & \vdots \\
A_*^{T-1}B_* & A_*^{T-2}B_* & \dots & B_*
\end{bmatrix}, \]

\[ \tilde{F} = \begin{bmatrix}
\tilde{A} & \tilde{B} \\
0 & I
\end{bmatrix}, \qquad \tilde{\eta} = \begin{bmatrix}
\eta_0 & \dots & \eta_{T-1} &  u_0 & \dots & u_{T-1}
\end{bmatrix}^\top \,, \]
and note that 
\[ \tilde{F} \tilde{\eta} = \begin{bmatrix}
x_1 & \dots & x_T & u_1 & \dots & u_T
\end{bmatrix}^\top \,. \]
Then, for any realization $\{\phi_k\}_{k=0}^{T}$,
\[
   \norm{\sum_{k=0}^{T-1} \phi_k \phi_k^\top} \leq \sum_{k=1}^{T} \phi_k^\top \phi_k = (\tilde{F} \tilde{\eta})^\top (\tilde{F} \tilde{\eta}) = \tr \left( \tilde{F} \tilde{\eta} \tilde{\eta}^\top \tilde{F}^\top \right) \,.
\]
We then note, letting $U = [u_0 \dots u_{T-1}]^\top [u_0 \dots u_{T-1}]$,
\[ 
\E[\tilde{\eta} \tilde{\eta}^\top] = \begin{bmatrix}
\sigma^2 I & 0 \\ 
0 & U
\end{bmatrix} \,.
\]
which implies
\begin{align}
    \E\left[ \norm{\sum_{k=0}^{T-1} \phi_k \phi_k^\top} \right] &\leq \tr\left( \tilde{F} \E\left[ \tilde{\eta} \tilde{\eta}^\top\right] \tilde{F}^\top\right) \\
    &\leq \tr\left( \begin{bmatrix}
    \sigma^2 \tilde{A} \tilde{A}^\top + \tilde{B}U\tilde{B}^\top & U \tilde{B}^\top \\
    \tilde{B}U & U
    \end{bmatrix}\right) \\
    &= \sigma^2 \tr\left( \tilde{A} \tilde{A}^\top\right) + \tr\left( \tilde{B}U\tilde{B}^\top \right) + \tr(U)
\end{align}
Letting $u_M = \max_{1 \leq i \leq T} u_i$, we then have
\[ \tr\left( \tilde{A} \tilde{A}^\top\right) = \sum_{k = 0}^{T-1} \tr(\Gamma_k (A_*)) \leq T \tr(\Gamma_{T-1}(A_*)) \]
\[ \tr\left(\tilde{B}U\tilde{B}^\top\right) \leq u_M^2 \sum_{k = 0}^{T-1}\tr(\Gamma_k (A_*, B_*)) \leq T u_M^2 \tr(\Gamma_{T-1} (A_*,B_*))  \]
Hence, by Markov's inequality,
\begin{align*}
&\P\left(\norm{\sum_{k=0}^{T-1} \phi_k \phi_k^\top} > \frac{\sigma^2 T \tr(\Gamma_{T-1}(A_*)) + T u_M^2 \tr(\Gamma_{T-1} (A_*,B_*)) + Tu_M^2}{\delta} \right) \\
&\qquad \leq   \P\left(\tr \left( \tilde{F} \tilde{\eta} \tilde{\eta}^\top \tilde{F}^\top \right)  > \frac{\sigma^2 T \tr(\Gamma_{T-1}(A_*)) + T u_M^2 \tr(\Gamma_{T-1} (A_*,B_*)) + Tu_M^2}{\delta} \right) \leq \delta
\end{align*}
That is, with probability at least $1 - \delta$, 
\begin{align*}
    Y_T = \sum_{k=0}^{T-1} \phi_k \phi_k^\top \preceq \left( \frac{\sigma^2 T \tr(\Gamma_{T-1}(A_*)) + T u_M^2 \tr(\Gamma_{T-1} (A_*,B_*)) + Tu_M^2}{\delta} \right) I \,.
\end{align*}
\end{proof}

\begin{proposition}[Upper Bound for Systems with Unmodeled Dynamics] \label{unmodeled-vup}
Consider the system in \eqref{unmodeled-intro-x}-\eqref{unmodeled-intro-w} where $w_k, \eta_k, x_k \in \R^n$, $\eta_k \sim \subG(\sigma^2)$, and $u_k \in \R^m$ are deterministic for each $k$. Define
\[ \phi_k = \begin{bmatrix}
x_k \\ u_k
\end{bmatrix}, \qquad Y_T = \sum_{k=0}^{T-1} \phi_k \phi_k^\top \,.\]
Then, for any $\delta > 0$, with probability at least $1 - \delta$,
\[  Y_T \preceq CT u_M^2 w_M^2 \sigma^2 \left( \Gamma(A_*) + \Gamma(A_*, B_*) \right)\bigg(1 + \log (1 / \delta) \bigg) I \,, \]
for a universal constant $C$.
\end{proposition}
\begin{proof}
We first define the quantities
\[ \tilde{A} = \begin{bmatrix}
I & 0 & \dots & 0 \\
A_* & I & \dots & 0 \\
\vdots & \vdots & \ddots & \vdots \\
A_*^{T-1} & A_*^{T-2} & \dots & I 
\end{bmatrix}, \qquad \tilde{B} = \begin{bmatrix}
B_* & 0 & \dots & 0 \\
A_*B_* & B_* & \dots & 0 \\
\vdots & \vdots & \ddots & \vdots \\
A_*^{T-1}B_* & A_*^{T-2}B_* & \dots & B_*
\end{bmatrix}, \]

\[ \tilde{F} = \begin{bmatrix}
\tilde{A} & \tilde{A} & \tilde{B} \\
0 & 0 & I
\end{bmatrix}, \qquad \tilde{\eta} = \begin{bmatrix}
\eta_0 & \dots & \eta_{T-1} & w_0 & \dots & w_{T-1} & u_0 & \dots & u_{T-1}
\end{bmatrix}^\top \,, \]
and note that 
\[ \tilde{F} \tilde{\eta} = \begin{bmatrix}
x_1 & \dots & x_T & u_1 & \dots & u_T
\end{bmatrix}^\top \,. \]
Then, again for any realization $\{\phi_k\}_{k=0}^{T}$,
\[
   \norm{\sum_{k=0}^{T-1} \phi_k \phi_k^\top} \leq \sum_{k=1}^T \phi_k^\top \phi_k = (\tilde{F} \tilde{\eta})^\top (\tilde{F} \tilde{\eta}) = \tr \left( \tilde{F} \tilde{\eta} \tilde{\eta}^\top \tilde{F}^\top \right) \,.
\]
For simplicity, write $u = [u_0 \dots u_{T-1}]^\top$, $w = [w_0 \dots w_{T-1}]^\top$, and $\eta = [\eta_0 \dots \eta_{T-1}]^\top$. Expanding the definition of $\tr \left( \tilde{F} \tilde{\eta} \tilde{\eta}^\top \tilde{F}^\top \right)$, we then see
\begin{align*}
    \tr \left( \tilde{F} \tilde{\eta} \tilde{\eta}^\top \tilde{F}^\top \right) &= \tr(\tilde{A} \eta \eta^\top \tilde{A}^\top) \\
    &\qquad + 2 \tr(\tilde{A} \eta w^\top \tilde{A}^\top)  + 2 \tr(\tilde{A} \eta u^\top \tilde{B}^\top) \\
    &\qquad + \tr(\tilde{A} w w^\top \tilde{A}^\top) + \tr(\tilde{B} u u^\top \tilde{B}^\top)
\end{align*}

We then see that $\tr(\tilde{A} \eta \eta^\top \tilde{A}^\top)$ can be bounded with high probability using \cite[Proposition 9.4]{Sarkar_2019}, and that $\tr(\tilde{A} \eta w^\top \tilde{A}^\top)$ and $\tr(\tilde{A} \eta u^\top \tilde{B}^\top)$ are sub-Gaussian random variables with variance proxy $T \Gamma(A_*) \sigma^2 w_M^2$ and $T\Gamma(A_*, B_*) \sigma^2 u_M^2$, respectively. Hence, with probability at least $1 - 3 \delta$, we will have
\begin{align*}
    \tr(\tilde{A} \eta \eta^\top \tilde{A}^\top) &\leq \sigma^2 \left( T \Gamma(A_*) \right) \left(1 + c_1 \log (1/\delta) \right) \\
    2 \tr(\tilde{A} \eta w^\top \tilde{A}^\top) &\leq \sqrt{c_2 T \Gamma(A_*) \sigma^2 w_M^2 \log (1/\delta)} \\
    2 \tr(\tilde{A} \eta u^\top \tilde{B}^\top) &\leq \sqrt{c_3 T \Gamma(A_*, B_*) \sigma^2 u_M^2 \log (1/\delta)} \,,
\end{align*}
where $u_M = \max_{0 \leq i \leq T} |u_i|$ and $w_M = \max_{0 \leq i \leq T} |w_i|$. We further note, by expanding definitions, that $\tr(\tilde{A} w w^\top \tilde{A}^\top) \leq T w_M \Gamma_T (A_*)$ and $\tr(\tilde{B} u u^\top \tilde{B}^\top) \leq T u_M \Gamma_T(A_*, B_*) $. Hence, with probability at least $1 - 3 \delta$, there is a universal constant $C$ such that
\[ Y_T \preceq \norm{\sum_{k=0}^{T-1} \phi_k \phi_k^\top}I  \preceq CT u_M w_M \sigma \left(  \Gamma(A_*) + \Gamma(A_*, B_*) \right)\bigg(1 + \log (1 / \delta) \bigg) I \,,\]
implying the claim above.
\end{proof}

\section{Proof of Lemma \ref{spectral-transfer}}
\label{a:transfer-proof}
We first note
\[ 
    \frac{1}{T}  \sum_{k = 0}^{T-1} y_k e^{-j2 \pi  \omega_0 k} = \frac{1}{T}  \sum_{k = 0}^{T-1} H(e^{j \omega_0}) u_k e^{-j2 \pi \omega_0 k} + \frac{1}{T}\sum_{k = 0}^{T-1}\eta_k e^{-j 2 \pi \omega_0 k} \,,
\]
by the definition of the Discrete Fourier Transform, since $\y(e^{j\omega}) = H(e^{j\omega}) \u(e^{j\omega}) + \boldeta (e^{j\omega})$ and $\omega_0 \in \Omega_T$. Rearranging, we have:
 \begin{align*}
    \frac{1}{T}  \sum_{k = 0}^{T-1} y_k e^{-j2 \pi \omega_0 k} - &H(e^{j \omega_0}) \bar{u} (\omega_0) =\\  &H(e^{j \omega_0}) \left( \frac{1}{T}  \sum_{k = 0}^{T-1}  u_k e^{-j2 \pi \omega_0 k}  - \bar{u} (\omega_0) \right)+ \frac{1}{T}\sum_{k = 0}^{T-1}\eta_k e^{-j2 \pi \omega_0 k} \,.
 \end{align*}
From Proposition \ref{sum-dependent}, the claim then follows immediately.

\section{Proof of Proposition \ref{spectral-to-PE}}
\label{a:s-to-pe}
Note that for any unit vector $z$, and any realization $\{\phi_k\}_{k=i}^{i+S}$,
\begin{align*}
z^\top \left( \frac{1}{S + 1}\sum_{k = i}^{i + S} \phi_k \phi_k^\top \right) z &=  \frac{1}{S + 1}\sum_{k = i}^{i + S} (\phi_k^\top z)^2 \\
&\geq \left\vert \frac{1}{S + 1}\sum_{k = i}^{i + S} \phi_k^\top z e^{-j  2\pi\omega k}\right\vert^2 \,,
\end{align*}
by Jensen's inequality. Then, we see
\begin{align*}
z^\top \left( \frac{1}{S + 1}\sum_{k = i}^{i + S} \phi_k \phi_k^\top \right) z &\geq \frac{1}{n} \sum_{\ell = 1}^n \left\vert \frac{1}{S + 1}\sum_{k = i}^{i + S} \phi_k^\top z e^{-j 2\pi \omega_\ell k}\right\vert^2 \\
&= \frac{1}{n} ||(\bar{\Phi} + W ) z||^2 \,.
\end{align*}
Here, $W$ is a random matrix for which each column is $R / (S + 1)$ sub-Gaussian. Continuing, we have:
\begin{align*}
z^\top \left( \frac{1}{S + 1}\sum_{k = i}^{i + S} \phi_k \phi_k^\top \right) z &\geq \frac{1}{n} \left( \norm{\bar{\Phi}^{-1}}^{-2} - ||W||^2 \right) \,.
\end{align*}
Finally, we see that because $W$ has (possibly dependent) entries which are $R/(S+1)$ sub-Gaussian, the claim follows from Proposition \ref{op-norm-dependent} setting $t = \norm{\bar{\Phi}^{-1}}^{-1}/2$.

% \section{Proof of Theorem \ref{system-id-thm}}
% \label{a:si-proof}
% \input{appendix/main-thm-proof.tex}

\section{Proof of Theorem \ref{unmodeled-estimation}}
\label{a:unmodeled-proof-upper}
We first rewrite \eqref{unmodeled-intro-x} as
\begin{align*}
\begin{bmatrix}
x_{k + 1} \\ u_{k + 1} 
\end{bmatrix} &= \begin{bmatrix}
A_* & B_* \\ 0 & 0 
\end{bmatrix} \begin{bmatrix}
x_k \\ u_k
\end{bmatrix} + \begin{bmatrix}
w_k + \eta_k \\ u_{k+1}
\end{bmatrix} \,,
\end{align*}
We can then see, considering this as a multi-dimensional regression problem, defining $\hat{F} = \begin{bmatrix}
\hat{A} & \hat{B} 
\end{bmatrix}$ and $F_* = \begin{bmatrix}
A_* & B_*
\end{bmatrix}$, that
\[ \hat{F} - F_* = \left((\Phi^\top \Phi)^\dagger \Phi^\top (W + E) \right)^\top \,,\]
where
\[ \Phi = \begin{bmatrix}
x_0^\top &  u_0^\top \\
\vdots & \vdots \\
x_{T-1}^\top & u_{T-1}^\top
\end{bmatrix}, \quad E = \begin{bmatrix}
\eta_0^\top \\ \vdots  \\ \eta_{T-1}^\top
\end{bmatrix} , \quad W = \begin{bmatrix}
w_0^\top \\ \vdots  \\ w_{T-1}^\top
\end{bmatrix} \,. \]

Defining the quantities
\[ Y_T = \Phi^\top \Phi, \quad S_T = \Phi^\top E \,, \]
we then see
\begin{equation} \label{unmodeled-split}
 \max \left\lbrace \norm{\hat{A} - A_*}, \norm{\hat{B} - B_*} \right\rbrace \leq \norm{(Y_T^{\dagger})^{1/2}}_2 \norm{(Y_T^{\dagger})^{1/2} S_T}_2  + \norm{Y_T^{\dagger}} \norm{\Phi^\top W}_2 \,.
\end{equation}

We may then define two events,

\begin{align}
    \cE_0 &= \{ V_{dn} \preceq Y_T \preceq V_{up}, T \geq T_0 \} \label{unmodeled-pe-bound} \\
    \cE_1 &= \left\lbrace \norm{S_T}_{(Y_T + V_{dn})^{-1}} \leq \sigma \sqrt{8(n+m) \log \left(\frac{5 \det(Y_T + V_{dn})^{1/(2(n+m))} \det(V_{dn})^{-1/2(n+m)} }{\delta^{1/(n+m)}} \right)} \right\rbrace \label{unmodeled-self-norm}
\end{align}

From Proposition \ref{spectral-to-PE} and Proposition \ref{unmodeled-vup}, we see that with probability at least $1 - 2\delta$, $\cE_0$ is satisfied with
\begin{align}
    V_{dn} &=  \frac{1}{2(n+m)} \norm{\Phi^{-1}}^{-2} T I \label{unmodeled-vdn-proof}\\
    V_{up} &= C T u_M^2 w_M^2 \sigma^2 \left( \Gamma_T(A_*) + \Gamma_T(A_*, B_*) \right)\bigg(1 + \log (1 / \delta) \bigg)I \label{unmodeled-vup-proof}\\
    T_0 &= \left(\log \frac{1}{\delta} + 2(n + m) \log 9\right) \frac{2(n+m) A_\omega \sigma^2}{c ||\bar{\Phi}^{-1}||^{-1}} \label{unmodeled-t}\,,
\end{align}
where $C$ is a universal constant and $A_\omega = \max_{i \in \Omega}\norm{(e^{j\omega_i}I - A_*)^{-1}}$ represents the maximum variance of the external disturbance in the frequency domain, with $\Omega$ defined as in Definition \ref{big-phi}.

Further, we see that by Proposition \ref{self-norm-body}, with probability at least $1 - \delta$, $\cE_1$ will hold for $V_{dn}$ as in \eqref{unmodeled-vdn-proof}. Combining these statements, we conclude that with probability at least $1 - 3 \delta$,
\begin{align*}
    \max \left\lbrace \norm{\hat{A} - A_*}, \norm{\hat{B} - B_*} \right\rbrace \overset{\eqref{unmodeled-split}}{\leq}& \norm{(Y_T^{\dagger})^{1/2}}_2 \norm{(Y_T^{\dagger})^{1/2} S_T}_2 \\ &\qquad + \norm{Y_T^{\dagger}} \norm{\Phi^\top W}_2 \\
    \overset{\eqref{unmodeled-pe-bound}, \eqref{unmodeled-self-norm}}{\leq}& \norm{V_{dn}^{-1/2}}_2 \sigma \sqrt{16(n+m) \log \left(\frac{5 \det(V_{up}V_{dn}^{-1} + I)^{1/2(n+m)}}{\delta^{1/(n+m)}} \right)} \\ 
    &\qquad + \norm{V_{dn}^{-1}} \norm{\Phi^\top W}_2 \,.
\end{align*}

What remains is to analyze the term $\norm{\Phi^\top W}_2$. We write it as follows:
\begin{align*}
     \norm{\Phi^\top W}_2 &= \norm{\begin{bmatrix}
     \sum_{k = 0}^{T-1} x_k w_k^\top \\
     \sum_{k = 0}^{T-1} u_k w_k^\top
     \end{bmatrix}}_2 \,. \\
     &= \norm{\begin{bmatrix}
     \sum_{k = 0}^{T-1} \phi_k w_k^\top
     \end{bmatrix}} \,,
\end{align*}
where $\phi_k = \begin{bmatrix} x_k^\top & u_k^\top \end{bmatrix}^\top$. We may then apply a multivariate version of Parseval's theorem, as
\begin{align*}
    \sum_{k = 0}^{T-1} \phi_k w_k^\top &= \sum_{k = 0}^{T-1} \phi_k \left( \frac{1}{T}\sum_{n = 0}^{T-1} \mathbf{w}(n) e^{j 2\pi kn/T} \right)^\top \\
    &= \sum_{k = 0}^{T-1} \phi_k \left( \frac{1}{T}\sum_{n = 0}^{T-1} \mathbf{w}(n) e^{-j 2\pi kn/T} \right)^\top &&(w_k \in \mathbb{R}^n, \forall k) \\
    &= \frac{1}{T} \sum_{k = 0}^{T-1} \sum_{n = 0}^{T-1} \phi_k \mathbf{w}(e^{jn/T})^\top  e^{-j 2\pi kn/T} \\
    &= \left(\frac{1}{T} \sum_{n = 0}^{T-1} \mathbf{w}(e^{jn/T}) \sum_{k = 0}^{T-1} \phi_k^\top  e^{-j 2\pi kn/T} \right)^\top \\
    &= \left(\frac{1}{T} \sum_{n = 0}^{T-1} \mathbf{w}(e^{jn/T}) \boldsymbol{\phi}(e^{jn/T})^\top \right)^\top \\
    &= \sum_{k = 0}^{T-1}  \frac{1}{T} \boldsymbol{\phi}(e^{jk/T}) \mathbf{w}(e^{jk/T})^\top \,. \numberthis \label{eq:parseval}
\end{align*}
Here, we recall that the bold-faced characters represent the DFT of each signal. 
From \eqref{eq:sg-spec-line}, we note that $\frac{1}{T} \boldsymbol{\phi}(e^{jk/T}) = \bar{\phi}(e^{jk/T}) + \epsilon_k$, where each $\epsilon_k$ is a sub-Gaussian random variable with variance proxy at most $\sigma^2 / T < \infty$.
Hence,
\begin{align*}
    \sum_{k = 0}^{T-1} \phi_k w_k^\top &= \sum_{k = 0}^{T-1}  (\bar{\phi}(e^{jk/T}) + \epsilon_k) \mathbf{w}(e^{jk/T})^\top \,.
\end{align*}
Given the definition of $\epsilon_k$, and applying the union bound, we see
\begin{align*}
    \P\left( \max_{1 \leq k \leq T} \|\epsilon_k \| \geq C\right) &\leq \sum_{k = 1}^T \P\left( \|\epsilon_k\| \geq C\right) \\
    &\leq 2T e^{-\frac{C^2 T}{2\sigma^2} } \,.
\end{align*}
Setting $C = \sqrt{\frac{2 \sigma^2 \log (2T / \delta)}{T} }$, we see that the probability that the maximal $\epsilon_k$ is large can be bounded by $\delta$.
Hence, with probability at least $1 - \delta$,
\begin{align*}
    \norm{\sum_{k = 0}^{T-1} \phi_k w_k^\top} &\leq \sum_{k = 0}^{T-1}  \norm{\bar{\phi}(e^{jk/T}) \mathbf{w}(e^{jk/T})^\top} + \tilde{O}\left( \frac{1}{\sqrt{T}} \sum_{k = 1}^T \| \mathbf{w}(e^{jk/T}) \| \right)\,.
\end{align*}
Putting the claims together, we then see with probability at least $1 - 4 \delta$,
\begin{align*}
    \max &\left\lbrace \norm{\hat{A} - A_*}, \norm{\hat{B} - B_*} \right\rbrace \\
    &\leq\norm{V_{dn}^{-1/2}}_2 \sigma \sqrt{16(n+m) \log \left(\frac{5 \det(V_{up}V_{dn}^{-1} + I)^{1/2(n+m)}}{\delta^{1/(n+m)}} \right)} \\ 
    &\qquad + \norm{V_{dn}^{-1}} \norm{\Phi^\top W}_2 \\
    &\leq \tilde{O}\left(\sqrt{\frac{1}{T \norm{\bar{\Phi}^{-1}}^{-2}} }\right) + \\ &\qquad \tilde{O}\left(\frac{1}{T \norm{\bar{\Phi}^{-1}}^{-2}}  \times \left(\sum_{k = 0}^{T-1}  (\bar{\phi}(e^{jk/T}) \mathbf{w}(e^{jk/T})^\top + 
    \tilde{O}\left( \frac{1}{\sqrt{T}} \sum_{k = 1}^T \| \mathbf{w}(e^{jk/T}) \| \right)\right)\right) \,,
\end{align*}
leading to the claim of the Theorem.

\section{Proof of Theorem \ref{lower-bound}}
\label{a:unmodeled-proof-lower}
Retaining the notation from the previous section, we recall
\[ \hat{F} - F_* = \left((\Phi^\top \Phi)^\dagger \Phi^\top (W + E) \right)^\top \,.\]
Moreover, we have
\begin{align*}
    2 \max \{ \norm{A - A_*}, \norm{B - B_*} \} &\geq \norm{A - A_*} + \norm{B - B_*} \\
    &\geq \norm{F - F_*} \,,
\end{align*}
since $\hat{F} - F_* = \begin{bmatrix} \hat{A} - A_* & \hat{B} - B_*\end{bmatrix}$.

We recall that $\norm{a + b} \geq \norm{a} - \norm{b}$, implying
\begin{align*}
    \norm{F - F_*}  \geq \norm{(\Phi^\top \Phi)^\dagger \Phi^\top W} - \norm{(\Phi^\top \Phi)^\dagger \Phi^\top E} \,,
\end{align*}
and from the previous section, we know that the events $\mathcal{E}_0$ and $\mathcal{E}_1$ can be defined to occur with probability at least $1 - \delta/2$ such that,
\[ \norm{(\Phi^\top \Phi)^\dagger \Phi^\top E} \leq \tilde{O}\left(\frac{1}{\sqrt{T}} \right) \,. \]
What remains is to lower bound the term $\norm{(\Phi^\top \Phi)^\dagger \Phi^\top W}$.
Under $\mathcal{E}_0$, and the event defined by \eqref{lower-bound-assumption}, we see
\begin{align*}
    \norm{(\Phi^\top \Phi)^\dagger \Phi^\top W} &\geq \norm{ V_{up}^\dagger \Phi^\top W} \\
    &\geq \frac{c}{T} \norm{\Phi^\top W} \\
    &\geq \frac{c}{T} \norm{\sum_{k = 0}^{T-1} u_k w_k^\top} \\
    &= \frac{c}{T} \norm{\sum_{k = 0}^{T-1} \frac{1}{T} \mathbf{u}(e^{jk/T}) \mathbf{w}(e^{jk/T})} \,,
\end{align*}
Where the final equality holds due to the multivariate extension of Parseval's theorem shown in \eqref{eq:parseval}.

Hence, if the assumption in \eqref{lower-bound-assumption} holds, with probability $1 - \delta/2$, $\norm{(\Phi^\top \Phi)^\dagger \Phi^\top W}$ is lower bounded by a constant which scales linearly with $\tau$.
Combining this with the previous calculations, we see that with probability $1 - \delta$,
\begin{align*}
    \max \{ \norm{A - A_*}, \norm{B - B_*} \} &\geq \frac{c}{2} \tau - \tilde{O}\left( \frac{1}{\sqrt{T}}\right) \,,
\end{align*}
completing the claim when $T$ is sufficiently large, on the order of $\tau^{-2}$.

\section{Proof of Theorem \ref{thm:regret}}
\label{a:regret-proof}
The proof of this claim follows directly from Theorem \ref{unmodeled-estimation} and Proposition \ref{prop:estimation-to-regret}.
We first denote the $i$th estimate of $(\hat{A}, \hat{B})$ as $(\hat{A^{(i)}}, \hat{B^{(i)}})$. 
Applying Theorem \ref{unmodeled-estimation} for each epoch, and applying a union bound with $\delta_i = O(1) \delta / (i+1)^2$, we then see that with probability $1 - \delta$, the following will hold for all epochs $i$:
\begin{equation} \label{eq:i-est}
    \max \{ \norm{\hat{A^{(i)}} - A_*},\norm{\hat{B^{(i)}} - B_*} \} \leq \frac{\bar{\epsilon} (A_*, B_*, Q, R)}{2^{i/4}} \,,
\end{equation}
where $\bar{\epsilon} (A_*, B_*, Q, R)$ is the constant specified in Proposition \ref{prop:estimation-to-regret}. This holds because the epoch length scales exponentially, as $T_i = C(A_*, B_*, Q, R, \delta) 2^i$, and because we may select the constant $C(A_*, B_*, Q, R, \delta)$ such that Theorem \ref{unmodeled-estimation} guarantees that after one epoch, the estimation error is at most $\bar{\epsilon} (A_*, B_*, Q, R)$ with probability at least $\delta_i$.
That is, because $\bar{\epsilon}(A_*, B_*, Q, R)^2$ depends only on system parameters and not the length of an epoch, so does $ C(A_*, B_*, Q, R, \delta)$.
Since the magnitude of the spectral lines-based exploration scales as $T_i^{-1/4}$, we then see that $\norm{\Bar{\Phi}^{-1}}^{-2} = c' 2^{-i/2}$ for each epoch $i$, so that \eqref{eq:i-est} holds.

From Proposition \ref{prop:estimation-to-regret}, this in turn implies that playing $u_k = K^{(i)} x_k$ in each epoch will result in 
\begin{align*}
    \hat{J^{(i)}} - J_* \leq \frac{\tau(A_*, B_*, Q, R) \bar{\epsilon} (A_*, B_*, Q, R)^2}{2^{i/2}} \,.
\end{align*}
To conclude, we note that since the amplitudes in Algorithm \eqref{alg:example} decay as $T_i^{-1/4} = 2^{-i/4}$, the total cost associated with the exploration of the spectral-lines based input will scale as $2^{-i/2}$, which is of the same order as the sub-optimality due to $K^{(i)}$.
Hence, we are able to show that
\begin{align*}
    \textrm{Regret}(T) &= \sum_{k = 0}^T (x_k^\top Q x_k + u_k^\top R u_k - J_*) \\
    &= \sum_{i = 1}^{O(\log_2 T)} \sum_k (x_k^\top Q x_k + (K^{(i)}x_k + u_{spec, k})^\top R (K^{(i)}x_k + u_{spec, k}) - J_*) \\
    &\overset{(*)}{\leq}  \sum_{i = 1}^{O(\log_2 T)} \sum_k \hat{J^{(i)}} - J_* +  x_{spec, k}^\top Q x_{spec, k} + u_{spec, k}^\top R u_{spec, k} \\
    &\leq \sum_{i = 1}^{O(\log_2 T)} \sum_k \tilde{O}(2^{-i/2}) \\
    &= \sum_{i = 1}^{O(\log_2 T)} \tilde{O}(2^{i/2}) \\
    &= \tilde{O} (\sqrt{T}) \,,
\end{align*}
completing the claim.
Here, the step in $(*)$ comes from the superposition principle, noting that the portion of $x_k$ derived from $u_k = K^{(i)}x_k + u_{spec, k}$ can be decomposed additively, such that $x_{spec, k}$ is the portion which depends on $u_{spec, k}$.

\section{Experiment Details}
\label{a:experiment-details}
In this section, we provide details and parameter sections for both experiments presented in Section \ref{s:experiments}.

\subsection{Parameter Estimation Errors}

In the experiments used to generate Figure \ref{fig:exp-results}, the following parameter selections are used for $A_*$ and $B_*$.

\begin{align}
    A_* = \begin{bmatrix}
    0 & 1 & 0 \\
    0 & 0 & 1 \\
    0.048 & -0.44 &  1.2
    \end{bmatrix}, \qquad B_* = \begin{bmatrix}
    0 \\ 0 \\ 1
    \end{bmatrix} \,.
\end{align}

In particular, $A_*$ satisfies Assumption \ref{assumption-stability} as it has eigenvalues $0.2, 0.4, $ and $0.6$. In the unmodeled dynamics \eqref{eq:unmodeled-nonlin}-\eqref{eq:unmodeled-highpass}, we set $\alpha = 0.001$, $c = 500$, and $\beta = 1$, in order to define the high pass filter which penalizes large deviations in $u_k - u_{k-1}$. Finally, the variance of the external noise, which is taken to be Gaussian, is set to be $\sigma^2 = 1$.

The frequencies $f_1$ and $f_2$ of the sinusoids were chosen as $f_1 = 0.01$ and $f_2 = 0.05$, as to be relatively low values which are still distinct.
Numerical optimization could also be performed at low frequencies as to optimize the bound in \eqref{system-id-thm}.

In order to perform the experiments, the system \eqref{unmodeled-intro-x}-\eqref{unmodeled-intro-w}, with the parameter selections above, is run for $T = 500$ iterations, and then this process is repeated $100$ times for each energy level of the control inputs. Parameter estimates are then computed using \eqref{least-squares} and the mean and standard deviation of the parameter estimation errors are presented in Figure \ref{fig:exp-results}.

\subsection{Regret of the Linear Quadratic Regulator Method}

The experiments used to generate Figures \ref{fig:regret-results} and \ref{fig:regret-results-unmodeled} used the following selection for $A_*$ and $B_*$:
\begin{align}
    A_* = \begin{bmatrix}
    0 & 1 & 0 \\
    0 & 0 & 1 \\
    1.03 & -3.06 &  3.03
    \end{bmatrix}, \qquad B_* = \begin{bmatrix}
    0 \\ 0 \\ 1
    \end{bmatrix} \,.
\end{align}
Here, $A_*$ has eigenvalues 0.9959, 1.01, and 1.0241, which imply that it is unstable, but we note that, as in line 1 of Algorithm 1, the initial controller is always stabilizing. The effective closed loop system matrix $A_* + B_* K^{(i)}$ is therefore always stable.

For the remaining system parameters, in the unmodeled dynamics \eqref{eq:unmodeled-nonlin}-\eqref{eq:unmodeled-highpass} we set $\alpha = 0.1$, $\beta = 0.9$, and $c = 4$ to simulate a high pass filter, and chose the noise-to-signal-ratio of the external system noise to be $0.01$ in the top plot of each figure and $0.1$ in the bottom plots.

The initial stabilitizing controller $K^{(0)}$ was given by a linear quadratic regulator applied to $B^*$ and an initial estimate of the dynamics matrix, $\hat{A}^{(0)}$. This was done to simulate the situation in which a system is being controlled with characteristics that are known to within some margin of error. $\hat{A}^{(0)}$ was generated by slightly perturbing the bottom row of $A^*$ with a Gaussian random variable.

\begin{figure}
    \renewcommand{\thefigure}{A}
    \centering
    \includegraphics[width=0.8\textwidth]{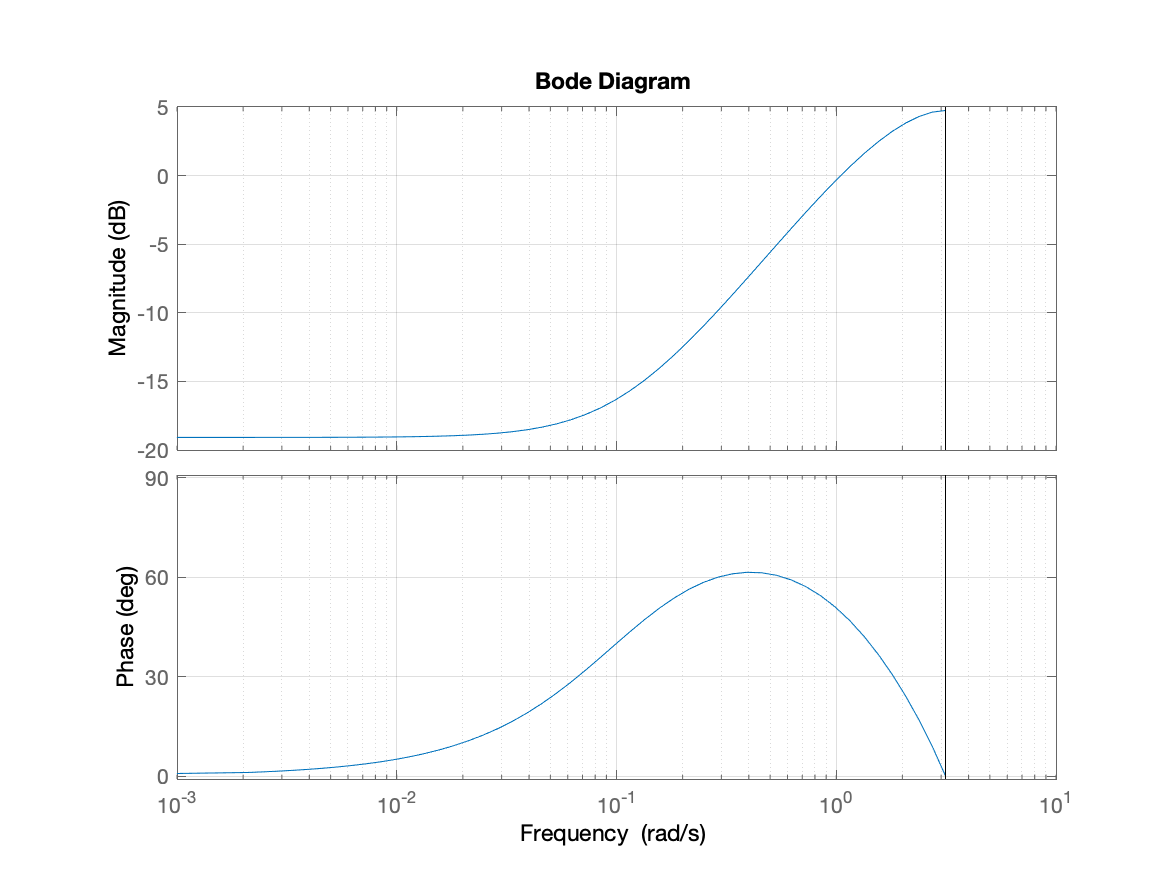}
    \caption{Bode plot for the unmodeled dynamics in regret experiments.}
    \label{fig:bode}
\end{figure}

In selecting the frequencies $f_1$ and $f_2$ for the spectral lines-based input for this problem, the Bode plot of the unmodeled dynamics (Figure \ref{fig:bode}) was used as a guide for frequency selection. In particular, we wanted frequencies at which the magnitude of the Bode plot was small so as to not excite the unmodeled dynamics: we selected $f_1 = 0.03$ and $f_2 = 0.05$.

We recall that the explicit definition of Regret used in Figures \ref{fig:regret-results} and \ref{fig:regret-results-unmodeled} can be defined as follows.
Given cost matrices $Q \succeq 0$ and $R \succeq 0$, we define an `optimal' cost:
\[ J_* = \min_u \lim_{T \rightarrow \infty} \frac{1}{T} \E\left[\sum_{k = 1}^T x_k^\top Q x_k + u_k^\top R u_k \right]  \qquad \text{s.t. dynamics } \eqref{lti-noisy-intro} \,.\]

Note here that this choice of $J_*$ uses the dynamics of the system \eqref{lti-noisy-intro}, primarily because the optimal solution may be computed using the discrete algebraic Ricatti equation. 
In general for systems which include unmodeled dynamics, a closed form for $J_*$ is not easily computed, and hence we use this proxy as the best achievable cost if there are no unmodeled dynamics. 
While $J_*$ as defined above is not a perfect optimal cost for regret purposes of the model \eqref{unmodeled-intro-x}-\eqref{unmodeled-intro-w}, in the sense that previous theoretical results in the literature may no longer hold, it still provides a reasonable baseline by which to compare the methods in experiments. 
The solution using the discrete algebraic Ricatti equation is shown as a baseline for all plots in Figures \ref{fig:regret-results} and \ref{fig:regret-results-unmodeled}, which allows for meaningful comparison to an `optimal' solution.

Figure \ref{fig:regret-results} shows the regrets of each approach in the scenario where there are no unmodeled dynamics. The system model used to produce Figure \ref{fig:regret-results} is then \eqref{lti-noisy-intro}, and the Regret$(T)$ used in that figure is defined as

\[ \text{Regret}_{Fig\_3}(T) = \sum_{k = 1}^T \left(x_k^\top Q x_k + u_k^\top R u_k - J_*\right) \text{s.t. dynamics } \eqref{lti-noisy-intro} \,.\]

However, Figure \ref{fig:regret-results} shows the regrets of each approach in the more realistic scenario where unmodeled dynamics are present. Since the true physical system is modeled according to \eqref{unmodeled-intro-x}-\eqref{unmodeled-intro-w}, the Regret$(T)$ used in Figure \ref{fig:regret-results-unmodeled} is then defined as

\[ \text{Regret}_{Fig\_4}(T) = \sum_{k = 1}^T \left(x_k^\top Q x_k + u_k^\top R u_k - J_*\right) \text{s.t. dynamics } \eqref{unmodeled-intro-x}-\eqref{unmodeled-intro-w} \,.\]

In the experiments, we set $Q = 10I$ and $R = 1$. To specify control inputs, we use Algorithm 1 as is for the ``Deterministic Sinusoidal Exploration'' and with $u_{spec, k}$ in line 6 replaced by Gaussian noise \eqref{eq:noise_input} for the ``Gaussian Noise Exploration.'' Experiments were run 50 times for 200 iterations, and the median and 90th percentile of Regret are shown in Figures \ref{fig:regret-results} and \ref{fig:regret-results-unmodeled}.

\subsection{Regret of LQR on Larger Systems}

The details of the simulations done on the larger system ($n = 5$ rather than $n = 3$) were identical except for the following changes. Figures \ref{fig:regret-results-larger} and \ref{fig:regret-results-unmodeled-larger} were produced with $A_*$ and $B_*$ selected as:
\begin{align}
    A_* = \begin{bmatrix}
    0 & 1 & 0 & 0 & 0 \\
    0 & 0 & 1 & 0 & 0 \\
    0 & 0 & 0 & 1 & 0 \\
    0 & 0 & 0 & 0 & 1 \\
    1.05 & -5.20 & 10.3 & -10.2 & 5.05
    \end{bmatrix}, \qquad B_* = \begin{bmatrix}
    0 \\ 0 \\ 0 \\ 0 \\ 1
    \end{bmatrix} \,.
\end{align}
Here, $A_*$ has eigenvalues 0.9959, 1.003, 1.01, 1.0171, and 1.0241. Additionally, the magnitude of the unmodeled dynamics was decreased to $c = 0.25$ and another frequency component ($f_3 = 0.07$) was added to the input to ensure persistent excitation.
\end{document}